\documentclass[a4paper,11pt]{article}
\usepackage{jcappub}
                     
\usepackage{amsmath}
\usepackage{blindtext, rotating}
\newcommand{\dv}[1]{\mathrm{d} #1}

\usepackage{orcidlink}
\graphicspath{{./}{figures/}}
\usepackage{array,colortbl}
\usepackage{multirow}

\usepackage[normalem]{ulem}
\usepackage{cleveref}
\usepackage{xspace}

\newcommand{\redmapper}{redMaPPer\xspace}

\newcommand{\degree}{\ensuremath{^{\circ}}\xspace}

\newcommand{\hMpc}{\ensuremath{\,h^{-1}\mathrm{Mpc}}\xspace}

\newcommand{\hMsun}{\ensuremath{\,h^{-1}M_\odot}\xspace}
\newcommand{\Msun}{\ensuremath{M_\odot}\xspace}
\newcommand{\Mbr}{$M_\mathrm{br}\,$}
\newcommand{\abr}{$\alpha_\mathrm{m}^\mathrm{br}\,$}

\newcommand{\rr}[1]{\textcolor{black}{#1}}
\newcommand{\Mstar}{{\ifmmode{\,M_{*}}\else{$M_{*}$}\fi}}
\newcommand{\Mhalo}{{\ifmmode{\,M_{\rm halo}}\else{$M_{\rm halo}$}\fi}}

\creflabelformat{equation}{#2\textup{#1}#3}

\title{The superclustering of hot gas: cosmological sensitivity in the Websky simulations}

\author[a]{M.~Lokken,\orcidlink{0000-0001-5917-955X}}
\affiliation[a]{Institut de F\'{i}sica d'Altes Energies (IFAE), The Barcelona Institute of Science and Technology, Campus UAB, 08193 Bellaterra (Barcelona) Spain}
\emailAdd{mlokken@ifae.es}

\author[b]{J.~R.~Bond,\orcidlink{0000-0003-2358-9949}}
\affiliation[b]{Canadian Institute for Theoretical Astrophysics, University of Toronto, 60 St. George St., Toronto, ON M5S 3H4, Canada}

\author[c,d]{R.~Hlo\v zek,\orcidlink{0000-0002-0965-7864}}
\affiliation[c]{David A. Dunlap Department of Astronomy \& Astrophysics, University of Toronto, 50 St. George St., Toronto, ON M5S 3H4, Canada}
\affiliation[d]{Dunlap Institute of Astronomy \& Astrophysics, 50 St. George St., Toronto, ON M5S 3H4, Canada}

\author[b,e]{N.~J.~Carlson,\orcidlink{0000-0002-2731-7708}}

\affiliation[e]{Department of Physics, University of Toronto, 60 St George St, Toronto, ON M5S 1A7, Canada}

\author[f]{Z.~Li,\orcidlink{0000-0002-0309-9750}}
\affiliation[f]{Berkeley Center for Cosmological Physics, 366 LeConte Hall, Berkeley, CA 94720, USA}

\author[g]{A.~van~Engelen\orcidlink{0000-0002-3495-158X}}
\affiliation[g]{School of Earth and Space Exploration, Arizona State University, Tempe, AZ 85287, USA}

\abstract{
Combinations of galaxy surveys and cosmic microwave background secondaries, such as the thermal Sunyaev Zel'dovich (tSZ) effect, are increasingly being used to jointly constrain cosmology and astrophysical properties of the gas within and beyond halos. Standard cross-correlations measure a directionless correlation between the microwave maps and galaxy catalogs. However, more information about the cosmic web structure can be captured by summary statistics which include environmental constraints and measure oriented correlations along axes of structure, such as filaments or superclusters. This work studies the sensitivity of multipole moments of constrained oriented stacks, a directional and environmentally-dependent statistic, to variations in cosmological and astrophysical parameters. We run nine different 2.4~Gpc-per-side simulations with the Websky algorithm, varying the dark matter energy density within flat $\Lambda$CDM, and create mock tSZ maps with each. We also apply six different gas prescriptions, imitating AGN feedback variations, to the fiducial cosmology. We analyze oriented stacks of the tSZ signal in supercluster regions in each simulation, focusing on signal out to $\sim20$ transverse Mpc from massive ($M>5\times10^{13}\Msun$) halos. The cosmology variations affect anisotropic and isotropic measurements similarly, while the halo-pasted gas variations mostly affect the isotropic signal. Our results suggest it is worthwhile to incorporate directional information into SZ-galaxy cross-correlations to increase cosmological sensitivity and help break degeneracies with gas physics.}

\begin{document}
\maketitle


\section{Introduction}\label{Sec:Intro} 
The superclustering of matter---the formation of anisotropic, nonlinear overdense structures spanning scales of a few to hundreds of megaparsecs---is a defining feature of the cosmic web. Anisotropic information, like the shapes and orientations of filaments and superclusters, contains imprints of the initial conditions of the universe \citep{vdWetal:2008b}. The evolution of these anisotropic features tells the story of how the dark matter-dominated tidal fields molded the pathways for gas to flow, channeling material into galaxies and galaxy clusters \citep{Cen1994, Frisch1995, Basilakos2001, Kolokotronis2002, Bharadwaj2004, Hopkins2005, Bagchi2017, Ho2018}.

Given the complexity of the cosmic web, the commonly-used isotropic statistics, such as two-point clustering as a function of the scalar $r$ (or power spectra of Fourier mode $k$), fall short: they only contain all of the information in the case of a Gaussian random field. In contrast, directionally-dependent statistics can capture distinct information about the non-Gaussian structure of the late Universe. In the current era of precision cosmology with rapidly incoming amounts of large-sky survey data, across a wide range of wavelengths, e.g. the Simons Observatory \citep{Zhilei2021};  the Dark Energy Spectroscopic Instrument \citep{DESI2016}; and the Vera Rubin Observatory's Legacy Survey of Space and Time \citep{LSST_Science_book2019}, it may be worthwhile to pursue directional measures that can be broadly applied to survey data. Such measurements may be necessary to identify subtle divergences from the concordance model $\Lambda$CDM (in which dark energy is the cosmological constant ($\Lambda$) and dark matter is collisionless and non-relativistic).

The superclustering of galaxies has been mapped through various methods \citep{melnyk2006, ForeroRomerp2009MNRAS.396.1815F, Disperse2013, Chon2013MNRAS.429.3272C, Tully2014, Chow-Mart2014, Bagchi2017, Kraljic2019MNRAS.483.3227K, PenarandaRivera2020}; however, the matter in galaxies represents only a small fraction \citep[$\sim$10\%,][]{read2005} of the baryonic matter in the universe. The rest is in various phases of gas in the intracluster medium, circumgalactic medium, and the diffuse intergalactic medium (IGM) \cite{Tumlinson2017, Rost2021}. The baryons beyond galaxies can be observed via scattering of photons from the cosmic microwave background (CMB) data, among other approaches.  Dominating the baryons by a factor of $\sim5$ is the dark matter; the combined matter distribution creates weak lensing distortions in galaxy and CMB observations. Thus, anisotropic superclustering statistics that can be applied to maps of CMB secondary effects are particularly valuable for understanding the distribution of the various cosmic web components. A statistical technique which combines projected maps and photometric or spectroscopic galaxy data, incorporates directionality, is flexible in scale, and allows for environmental constraints was presented in \citep[][hereafter L22]{Lokken2022PaperI}. This \textit{constrained oriented stacking} technique, detailed in \Cref{sec:methods}, provides cross-correlational shape information at constrained regions of the cosmic web. It has been applied to data from the Atacama Cosmology Telescope \citep[ACT,][]{ACT2016} and Dark Energy Survey \citep[DES,][]{DES2005} to demonstrate evidence at up to 10$\sigma$ for large-scale thermal energy anisotropy in supercluster regions \citep{Lokken2022PaperI, Lokken2025ApJ...982..186L}. These measurements have been compared directly to mock data from simulations; however, the mocks were limited to a single cosmology and only two distinct gas physics models.

In this work, we seek to demonstrate the sensitivity of oriented tSZ stacks, and their associated summary statistics, to cosmological parameter variations within the $\Lambda$CDM model. We use the Websky simulations and underlying Peak Patch algorithm \citep{BondMyers1996, Stein2019, Stein2020} to produce nine light-cone simulations of one-eighth of the sky extending to $z=0.65$ in redshift, each with a different set of cosmological parameters. The goal is to determine whether anisotropic information can add power to distinguish between universes with distinct energy densities of dark energy ($\Omega_\Lambda$) and matter ($\Omega_\mathrm{M}$).

Motivated by the previous and ongoing efforts to map the superclustering of gas using CMB secondary effects, we post-process the simulations to create mock maps of the tSZ effect \cite{SZ1970, SZ1972}, parametrized by Compton-$y$ \citep[see][for a review]{Mroczkowski2019}. This frequency-dependent effect is sensitive to inverse Compton scatterings of cold CMB photons off of hot ionized gas along the line-of-sight. It is related to the integrated gas pressure over the line-of-sight, and therefore is a tracer of the cosmic web (especially in hot and dense regions) and a probe of the thermal energy history of the universe \citep{Chiang2020ApJ...902...56C, Chen2023ApJ...953..188C, DiMascolo2024arXiv240300909D}. Cosmological inference from the tSZ effect is limited by our current uncertainties in the distribution of baryons, as studied in \cite{KomatsuSeljak2002, Bhattacharya2012, hill2013, Hill2013PhRvD..87b3527H, Hojjati2017MNRAS.471.1565H, Hill2018}. Thus, we also test the sensitivity of the oriented superclustering signal to variations in gas modeling by creating mock maps with different pressure profile prescriptions. 

This paper is structured as follows. We begin with an overview of the simulations in \Cref{sec:simulations}. \Cref{sec:methods} describes the oriented stacking methodology. \Cref{sec:cosmology_analysis} studies the sensitivity to cosmology. \Cref{sec:gas_pasting} describes the variations in gas models pasted onto halos and analyzes the resulting maps. We discuss the results and conclude in \Cref{sec:conclusion}.

\section{Websky cosmological simulations} \label{sec:simulations}
We use the mass-Peak Patch algorithm and associated Websky map-making algorithm \citep[hereafter referred to jointly as Websky,][]{BondMyers1996, Stein2019, Stein2020} to produce mock galaxy density- and Compton-$y$ maps under varied cosmologies. Although cosmological hydrodynamic simulations provide the most physically-motivated predictions of the evolution of the large-scale gas and galaxy distributions, they are computationally expensive to run. Even the largest existing volumes are challenging to map directly onto the enormous lightcone coverage of modern surveys, and also require intensive post-processing to convert from physical quantities to observable signals. Websky rapidly produces lightcone halo catalogues, using second-order Lagrangian perturbation theory and equations for ellipsoidal collapse to evolve each halo to a precise redshift (thus avoiding the need to stitch together redshift shells). The algorithm is orders-of-magnitude faster than an $N$-body simulation for the same volume. It has been validated against $N$-body codes (e.g., Figs. 7 and 8 of \cite{Stein2019}) and performs well for the quasi-linear scales that will be the focus of our analysis. It also compares favorably with other approximate methods for producing mock halo catalogues, such as \textsc{Cola} \cite{Lippich2019MNRAS.482.1786L, Blot2019MNRAS.485.2806B, Colavincenzo2019MNRAS.482.4883C}. The associated Websky algorithms \citep{Stein2020} contain infrastructure for converting the halo catalogues to maps of of CMB secondary effects including tSZ, kinematic SZ, cosmic infrared background, and CMB lensing convergence $\kappa$. In the case of tSZ, this involves pasting gas pressure profiles to create a mock $y$ map. The resulting $y$ power spectra have been compared to data in \cite{Stein2020}, showing accuracy across a broad range of scales but small-scale deviations. Improving the accuracy of the gas painting, such as for relativistic halos and small scales, is the subject of ongoing work \cite[e.g.,][]{Kuhn2025arXiv250418637K}. The speed of Websky, in combination with its large-scale accuracy, makes it an apt choice for this work: we can rapidly generate maps with large sky footprints (mimicking wide-field survey data) under several cosmological and gas parameter variations.

In practice, the user inputs a $z=0$ linear matter power spectrum, from which the algorithm generates an initial (Lagrangian-space) density field. For a lightcone (evolving) run, for each redshift interval the algorithm scans for peaks at varying filter scales and selects the most massive peak that will collapse to a halo by that redshift. The peak regions undergo homogeneous ellipsoid collapse to form halos, and an exclusion algorithm removes overlapping halos in order to conserve mass \citep{BondMyers1996, Stein2019}. The halos are moved to their final positions via second-order Lagrangian perturbation theory. The outputs include a halo catalog with positions and masses, and maps of the desired CMB secondaries.

\subsection[Varying LCDM parameters]{Varying $\Lambda$CDM parameters}

To test the sensitivity of oriented tSZ stacks to cosmology, we vary the cosmological parameters (and, accordingly, the linear matter power spectrum) input to the Peak Patch algorithm. We fix the seed for generating the initial density field such that structure grows in the same locations across the simulations. For each variation we create mock maps of the associated halo mass density and tSZ fields using the standard pressure profile implemented in Websky \cite{Battaglia2012b}.

We set the fiducial cosmology to the baseline $\Lambda$CDM model from \textit{Planck} 2018\footnote{Specifically we use the `TT,TE,EE+lowE+lensing' base-$\Lambda$CDM model specified in Table 2 of the \textit{Planck} paper.} \citep{Planck2018}. In this cosmology, the energy density of matter $\Omega_{\mathrm{M}}=0.3153$, the curvature energy density $\Omega_k=0$, the cosmological constant density $\Omega_\Lambda=0.6847$, the Hubble constant $H_0= 67.36\,\mathrm{km}\, \mathrm{s^{-1}}\, \mathrm{Mpc^{-1}}$, the scalar spectral index $n_s=0.9649$, and there are two massless neutrino species and a massive species with mass $m_\nu = 0.06$ eV. For variations around the fiducial model, we choose two simple sets of parameter trajectories. Of course, the full possible parameter space is large, and choosing well-motivated samples is an important part of inference from data. In our case, we do not attempt to do inference; we aim to simply explore how the dark matter-dark energy balance impacts the superclustering of hot gas, which is relatively uncharted territory. The trajectories we choose, therefore, cover only a small part of the parameter space to serve as a proof-of-concept.

The two trajectories correspond to running two suites of simulations, each including four variations around the fiducial \textit{Planck} cosmology (nine simulations in total). In both, we vary $\Omega_{\mathrm{M}}$ around the fiducial value such that $\Omega_{\mathrm{M}} = (0.22, 0.27, 0.32, 0.37, 0.42)$. To maintain a flat universe, we also vary $\Omega_\Lambda$ such that $\Omega_\mathrm{M}+\Omega_{\Lambda}=1$; therefore $\Omega_{\Lambda}$ takes the corresponding values $(0.78, 0.73, 0.68, 0.63, 0.58)$. The specifics of the two suites are shown in Table~\ref{tab:cosmo_var}, with motivation and details described in following sections.

The consequences of increasing $\Omega_{\mathrm{M}}$ include an earlier transition from the radiation- to matter-dominated eras, while later in the universe there is a delay in the transition to dark energy domination. (The opposite is true for decreasing $\Omega_\mathrm{M}$.) The change in matter-radiation equality impacts the position of the matter power spectrum turnover, and the delayed onset of dark energy allows for larger structures to collapse at late times. Given the many impacts, the value of $\Omega_{\mathrm{M}}$ in a $\Lambda$CDM universe is highly constrained; current mean and $68\%$ limits from \textit{Planck} are $\Omega_{\mathrm{M}} = 0.317^{+0.017}_{-0.016}$  for the `base plikHM TTTEEE lowl lowE' configuration \citep{Planck2018} and $\Omega_{\mathrm{M}} = 0.339^{+0.032}_{-0.031}$ from the DES Year 3 galaxy clustering and weak lensing constraints \citep{Abbott2022PhRvD.105b3520A}. Our variations go well beyond this range to clearly demonstrate the impacts of cosmology on superclustering; we leave inference of constraints from data to future work.

\begin{table*}[!htpb]
    \centering
    \begin{tabular}{c|c|c}
        Parameter & Value(s) in Cosmo1 & Value(s) in Cosmo2 \\ 
        \hline
        $\Omega_{\mathrm{M}}$  & (0.22, 0.27, 0.32, 0.37, 0.42) & (0.22, 0.27, 0.32, 0.37, 0.42) \\
        $\Omega_\Lambda$ & (0.78, 0.73, 0.68, 0.63, 0.58) & (0.78, 0.73, 0.68, 0.63, 0.59) \\
        $\Omega_{\mathrm{b}}$ & 0.049 & (0.034, 0.041, 0.049, 0.057, 0.065) \\
        $f_b$ & (0.23 0.19 0.16 0.13 0.12) & 0.16\\
        $\Omega_{\mathrm{M}} h^2$ & (0.097, 0.120, 0.143, 0.166, 0.188) & 0.143 \\
        $\Omega_{\mathrm{b}} h^2$ & 0.022 &  0.022\\
        $\Omega_\Lambda h^2$ & (0.36, 0.33, 0.31, 0.29, 0.27) & (0.52, 0.40, 0.31, 0.25, 0.20) \\
        $H_0$ & 67 & (82, 73, 67, 63, 59) \\
        $\sigma_8(z=0.5)$ & 0.62 & 0.62 \\
        $A_s \cdot 10^9$ & (1.93, 2.02, 2.10, 2.17, 2.24) & (1.94, 2.02, 2.10, 2.17, 2.24) \\
        \hline
        \hline
        $z_{\rm{m-\Lambda}}$ & (0.54, 0.40, 0.29, 0.20, 0.12) & (0.54, 0.40, 0.29, 0.20, 0.12)\\
        $\chi(z=0.5)$ [Mpc] & (2024, 1987, 1952, 1920, 1890) & (1672, 1822, 1952, 2066, 2168) \\
        age($z=0.5$) [Gyr] & (9.9, 9.2, 8.6, 8.1, 7.7) & (8.2, 8.4, 8.6, 8.7, 8.8)
    \end{tabular}
    \caption{Parameter variations for the Cosmo1 and Cosmo2 suites. Below the double lines, several derived properties are shown: the redshift of matter-lambda equality as well as the comoving distance $\chi$ and age of universe, both evaluated at $z=0.5$.}
    \label{tab:cosmo_var}
\end{table*}

\subsection[Fixing sigma-8 at z=0.5]{Fixing $\sigma_8$ at $z=0.5$}
Varying $\Omega_{\mathrm{M}}$ with a fixed primordial power spectrum amplitude $A_s$ would change the amplitude of matter fluctuations at late times. This amplitude is typically parametrized by $\sigma_8(z)$, the root-mean-square fluctuation of the linear matter overdensity field in spheres of $8\hMpc$ radius at redshift $z$. It is well understood that $\sigma_8$ is the predominant factor affecting the tSZ power spectrum \citep{Komatsu1999, KomatsuSeljak2002, Bolliet2018MNRAS.477.4957B}, thus we expect it to largely determine the strength of the stacked tSZ signal. Our focus, however, is on the sensitivity of the anisotropic tSZ signal to differences in structure evolution arising from a varied matter-dark energy balance. To isolate these more subtle effects from the rescaling of the amplitude of fluctuations, we adjust the simulation configuration with the goal of equalizing the power normalization across the variations at the redshift of analysis. In other words, to remove the $\sigma_8$ dependency, we vary $A_s$ such that $\sigma_8(z^*)$ is fixed to a constant value across all simulations, where $z^*=0.5$ is the redshift we will analyze.

We choose $z^*=0.5$ because it lies far enough away that a narrow interval ($\Delta z\sim0.02$) covers a cosmological volume containing a large number of clusters for stacking, while also being late enough for $\Lambda$ to have influenced structure formation in most models. The redshift of matter-$\Lambda$ equality $z_{\rm m-\Lambda}$ ranges from $0.12<z<0.54$ across the models, such that in the high-$\Omega_{\mathrm{M}}$ models dark energy has had little impact by $z=0.5$.

\begin{figure}
    \centering
    \includegraphics[width=.5\columnwidth]{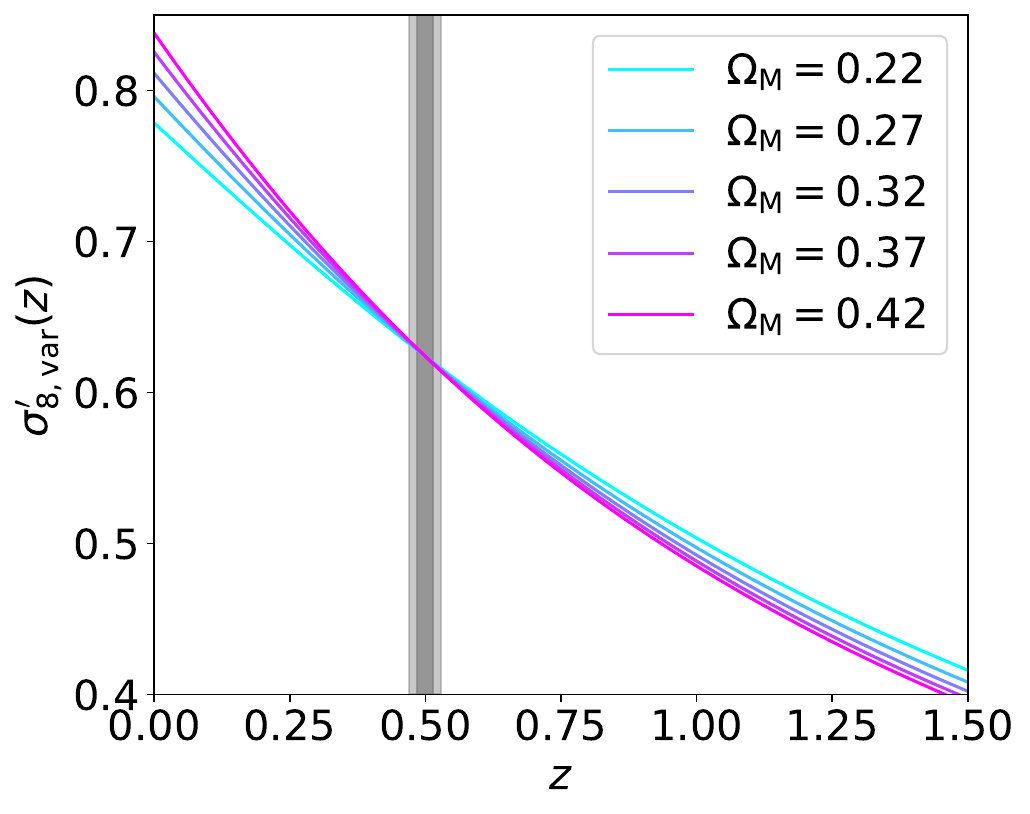}
    \caption{$\sigma_\mathrm{8,var}(z)'$, as defined in Eq.~\ref{eq:sigma8_var}, for the simulated cosmologies. Each cosmology has a different growth history, but the initial power spectrum normalization has been chosen such that $\sigma_\mathrm{8}$ converges at $z=0.5$ across all simulations. The narrower, darker band spans the range from which we select the stacking halo sample ($\sim100$~Mpc) and the wider band shows the range over which halos are projected into 2D number density maps, used for environmental selections and orientations ($\sim200$~Mpc).}
    \label{fig:sigma8_z}
\end{figure}
In the fiducial cosmology,  $\sigma_{8,\mathrm{fid}}(z^*)=0.62$. For a non-fiducial simulation (denoted by `var' in the following equations), we rescale the power spectrum at $z^*$ to enforce that  $\sigma'_{8,\rm{var}}(z^*)=0.62$, where the prime (') denotes the rescaled fluctuations. To do so, we first generate the linear power spectrum $P_{\rm{var}}(k, z^*)$ using \texttt{CAMB} \citep{CAMB2000}. Next, we renormalize such that:
\begin{equation}
    P_{\mathrm{var}}'(k, z^*) = P_{\mathrm{var}}(k,z^*)  \Big(\frac{\sigma_{8,\mathrm{fid}}(z^*)}{\sigma_{8,\mathrm{var}}(z^*)}\Big)^2.
\end{equation}
To retrieve $P'_\mathrm{var}(k, z=0)$, the required input to Peak Patch, we grow the power from $z=0.5$ to $z=0$ using the growth factor:
\begin{align} \label{eq:growth_factor}
    D(a) &= D_+(a)/D_+(a=1); \\
    D_+(a) &= \frac{5\Omega_{\mathrm{M}}}{2} \frac{H(a)}{H_0}\int_{0}^{a} \frac{da'}{(a'H(a')/H_0)^3},
\end{align}
where $a$ is the scale factor, $a=1/(1+z)$. Specifically, for the growth to $z=0$,
\begin{equation}
    P_{\mathrm{var}}'(k, z=0) =\frac{P_{\mathrm{var}}'(k, z^*)D_{\rm{var}}^2(z=0)} {D_{\rm{var}}^2(z^*)}.
\end{equation}
We use the \texttt{Colossus} cosmology package\footnote{\url{https://bdiemer.bitbucket.io/colossus/index.html}} for all calculations involving $\sigma_8$ and $D$.
Note that the rescaling procedure is equivalent to varying $A_s$ across the simulations such that $A_{s,\mathrm{var}}$ for each simulation is scaled by the factor $\Big(\frac{\sigma_{8,\mathrm{fid}}(z^*)}{\sigma_{8,\mathrm{var}}(z^*)}\Big)^2$.

The relationship between fluctuations in the rescaled vs original cosmology is given by:
\begin{equation}
    \sigma'_{8,\mathrm{var}}(z) = \frac{\sigma_{8,\mathrm{fid}}(z^*) D_{\mathrm{var}}(z)}{D_{\mathrm{var}}(z^*)}.
    \label{eq:sigma8_var}
\end{equation}
\Cref{fig:sigma8_z} shows how, after rescaling, all models pass through $\sigma_{8,\mathrm{fid}}$ at $z=0.5$, while the linear variance evolves around that point according to the parameters of that cosmology.

\subsection{The Cosmo1 variations}

In the first suite, henceforth called `Cosmo1', we vary $\Omega_\mathrm{M}$ as described above, but fix all other independent parameters to the fiducial model, including the baryon density $\Omega_{\mathrm{b}}$ and the Hubble constant $H_0$. $\Omega_{\mathrm{M}}$ propagates into the matter field at any given time as $\overline{\rho}_m(a) = \Omega_{\mathrm{M}} a^{-3} \rho_{\rm{cr}}$, where $\rho_{\rm{cr}}=\frac{3H_0^2}{8\pi G}$. Therefore, because $H_0$ is fixed, the $\Omega_{\mathrm{M}}$ variations impact the mean matter density at all redshifts. Further, because $\Omega_{\mathrm{b}}$ is fixed while $\Omega_{\mathrm{M}}$ varies, the cosmic baryon fraction ($f_b \equiv \Omega_\mathrm{b}/\Omega_\mathrm{M}$) varies. While $\Omega_\mathrm{M}$ impacts the power spectrum and thus the halo mass function (HMF), $f_b$ impacts the cluster pressure profiles: at a given mass, clusters in a lower-$\Omega_{\mathrm{M}}$ simulation have higher pressure, because $f_b$ is larger when $\Omega_\mathrm{b}$ is held fixed. Finally, varying $\Omega_\mathrm{M}$ and $\Omega_\Lambda$ changes the age of the universe at $z=0.5$ such that the universe is younger in the higher-$\Omega_\mathrm{M}$ runs, therefore less evolution has occurred up to this redshift (and vice versa for the lower-$\Omega_\mathrm{M}$ runs.) All Cosmo1 parameters are listed in the left column of Table~\ref{tab:cosmo_var}.

\subsection{The Cosmo2 variations}
The physical total matter density $\Omega_{\mathrm{M}}h^2$ and baryonic matter density $\Omega_{\mathrm{b}}h^2$, where $h=H_0/(100~\mathrm{km~s^{-1}~Mpc^{-1}}$), are the quantities better-constrained by CMB data, rather than $\Omega_{\mathrm{M}}$ and $\Omega_{\mathrm{b}}$ alone. Motivated by this, we explore another set of cosmological variations that encodes variation in the $\Omega_{\mathrm{M}}-\Omega_{\Lambda}$ plane while fixing the physical mean matter densities. This requires a change in the critical density of the universe and thus variations to $H_0$, rescaling the size of the universe. Specifically, for this `Cosmo2' suite, we vary $\Omega_{\mathrm{M}}$ across the same range as Cosmo1, but simultaneously vary $h$ such that $\Omega_{\mathrm{M}}h^2$ and $\Omega_{\mathrm{b}}h^2$ are fixed to their \textit{Planck} 2018 values \citep{Planck2018}. We perform the same fixing of $\sigma_8(z=0.5)$ on these runs. It is worth noting that this fixes the variance at $8\hMpc$, not 8~Mpc, across the variations; thus due to varying $H_0$, the comoving scale at which the variance is fixed is different across the Cosmo2 simulations. However, we conduct our analysis in angular sky coordinates, and the factor of $h$ cancels out in the conversion from transverse comoving coordinates to angles \citep{Hogg1999astro.ph..5116H}. As in the Cosmo1 suite, the age of the universe also varies across Cosmo2, but in a different manner due to varying $H_0$: the lower-$\Omega_\mathrm{M}$ simulations probe slightly younger universes than the higher-$\Omega_\mathrm{M}$ runs.  All Cosmo2 parameters are listed in the right column of Table~\ref{tab:cosmo_var}.

In terms of the effects on the pressure profile prescription, in Cosmo2 $f_b$ is fixed, but the variations in $H(z)$ are expected to propagate into the pressure profile prescription, as $\rho_{\rm{cr}}$ scales the amplitude of the pressure profile for clusters at a given $M$ and $z$ (see \Cref{sec:gas_pasting}). This dependence arises through the arguments of self-similarity \citep{Kaiser1986} under which the cluster gas prescription can be applied, with the $f_b$ and $\rho_{\rm{cr}}$ scalings, across differing cosmologies.

\subsection{A quick look at the simulations}
The cosmological variations are summarized in Table~\ref{tab:cosmo_var}. For each parameter set, we run a single lightcone realization for $1/8^\mathrm{th}$ of the sky extending to $z=0.65$ (2400~Mpc-per-side boxes). Resolution in Peak Patch is set by the user-defined grid size, related to the smallest filter scale used to find peaks in the Lagrangian density field. For computational speed, we choose to set the grid at a scale to produce $5\times10^{12}\Msun$ halos at minimum. Near this lower limit, we find that numerical limitations of the simulation cause the HMF to taper artificially toward lower masses, under-producing small halos. We thus restrict the resulting halo catalogues to halos with $M>10^{13}\Msun$, well above the resolution limit, to ensure that only halos in the realistic part of the HMF are included.

\begin{figure*}
    \centering
    \includegraphics[width=.45\textwidth]{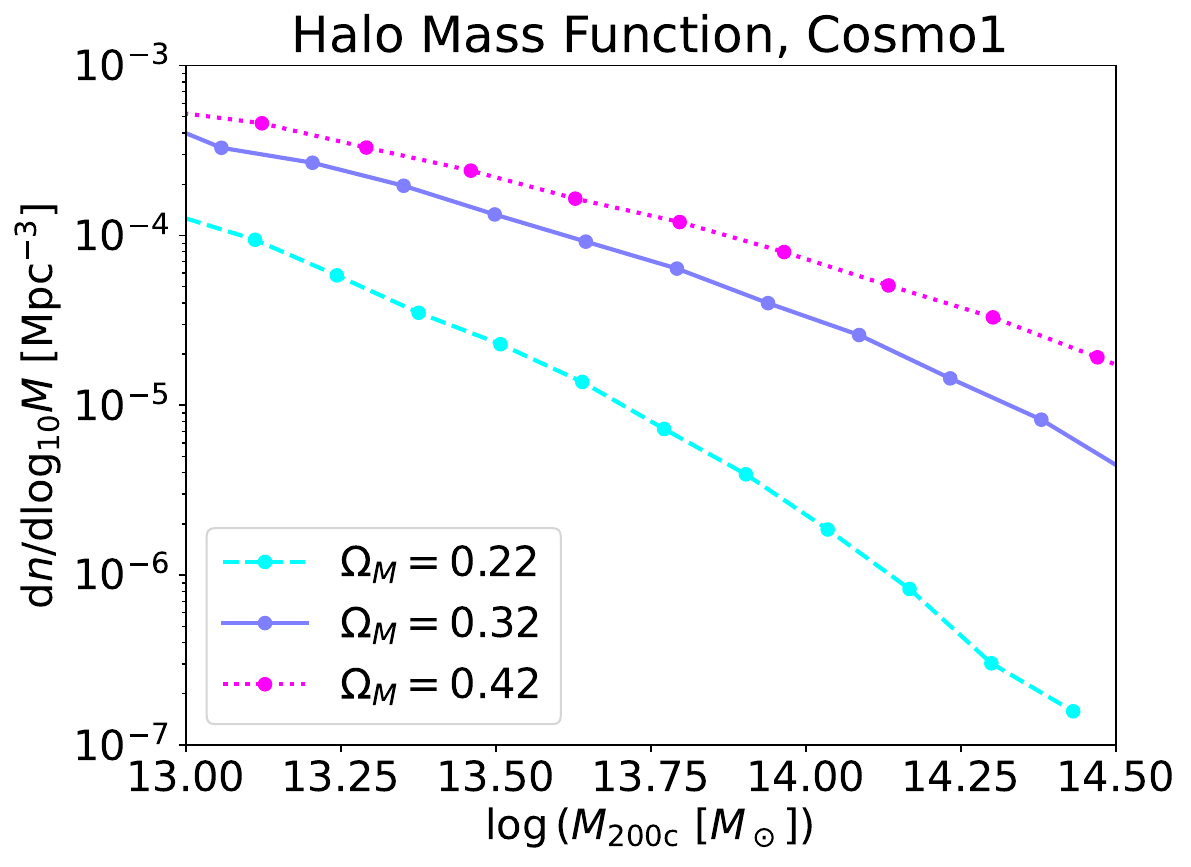}
    \includegraphics[width=.45\textwidth]{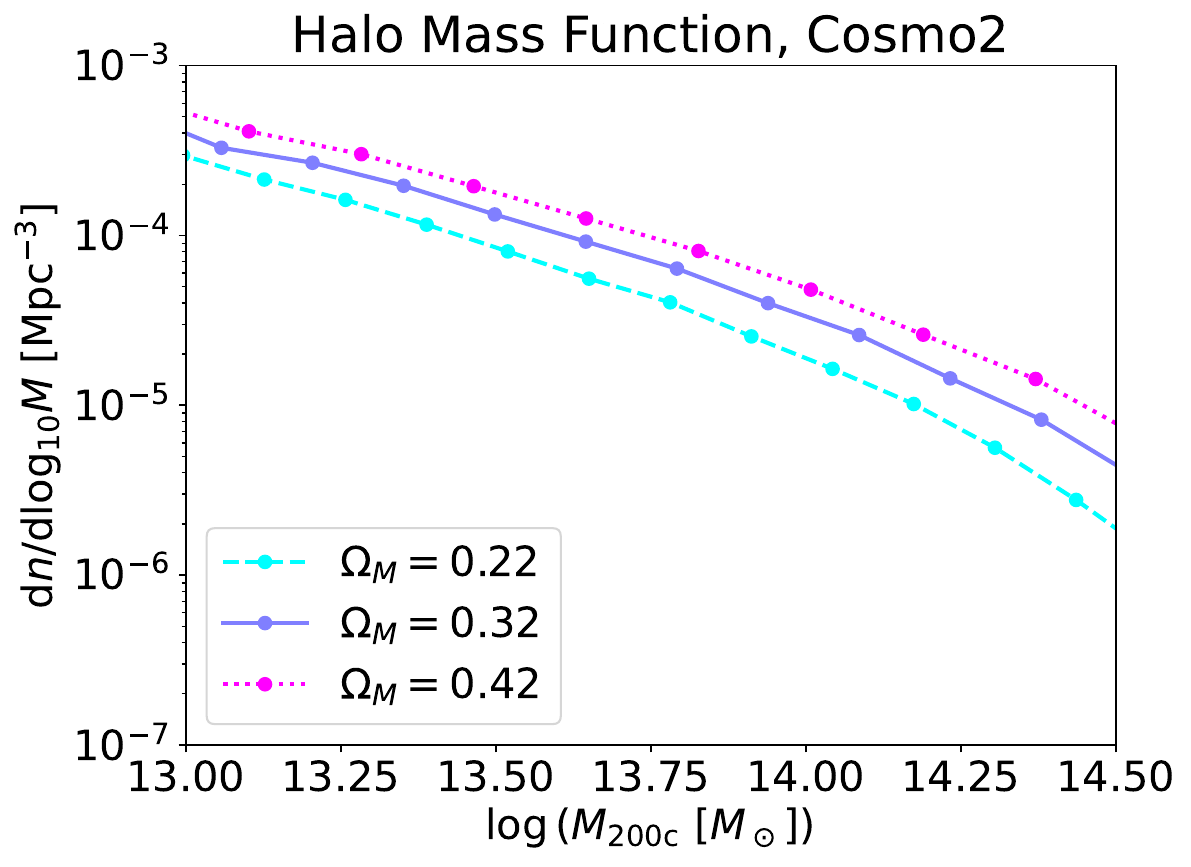}
    \includegraphics[width=\textwidth, trim={0 3cm 0 0},clip]{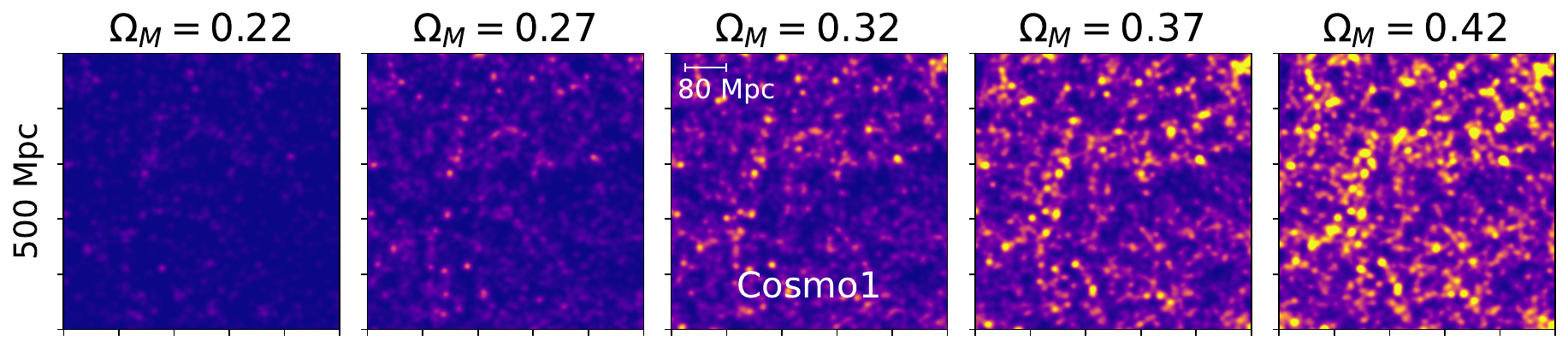}
    \includegraphics[width=\textwidth, trim={0 0 0 1cm},clip]{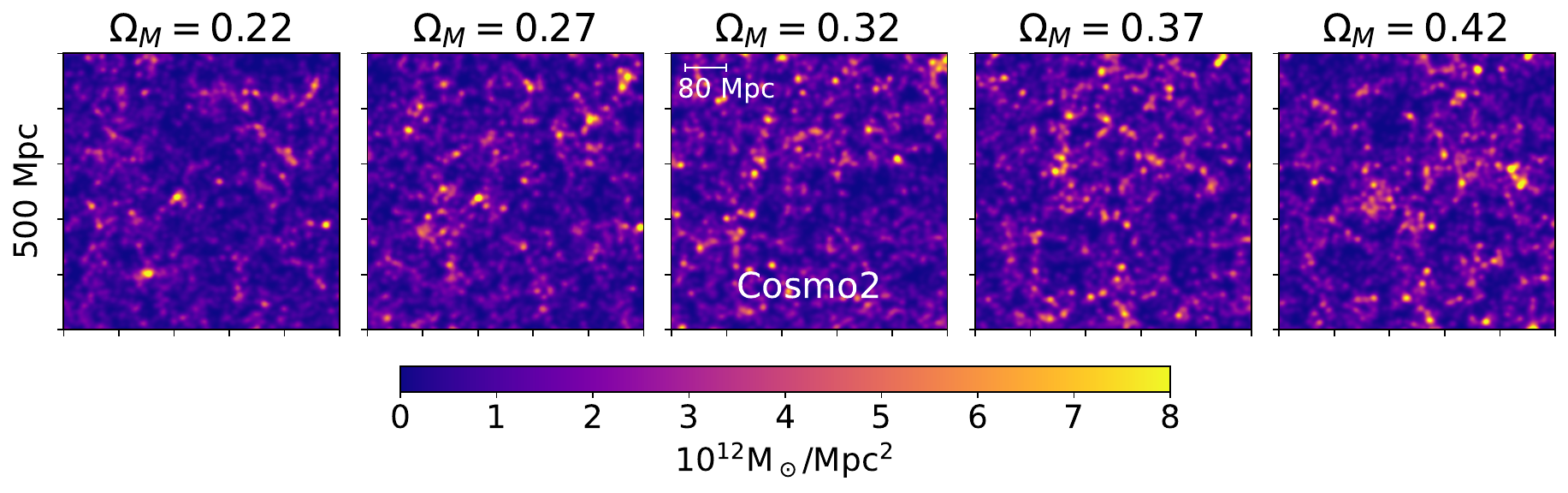}\\
    \caption{\textbf{Above}: halo mass functions for Cosmo1 (\textbf{left}) and Cosmo2 (\textbf{right}) at $z=0.5$, over the mass range used. Cosmo2 exhibits smaller differences due to the fixed matter density $\Omega_{\mathrm{M}}$. \textbf{Below}: halo surface mass density maps, smoothed with a Gaussian filter with FWHM=10 Mpc (half of the smoothing scale used for orientation), for all variations in Cosmo1 (\textbf{upper}) and Cosmo2 (\textbf{lower}). All simulations are run with the same initial seed. All plots cover a sky region of $\sim15^{\circ}\times15^{\circ}$ (which is 500~Mpc per side in the fiducial cosmology) projected along the $200$~Mpc line-of-sight axis, centered at $z=0.5$. The data is from a lightcone run, so within the $200$~Mpc extent, there is some evolution. The 80~Mpc scale bar corresponds to the side-length of the fiducial cosmology image in \Cref{fig:omegamh2fix_stack_img}. The comoving distance to $z=0.5$ depends on cosmology, so within adjacent Cosmo1 plots, the halos are extracted from slabs that are $\sim$30--40 Mpc offset in comoving line-of-sight distance. Thus, the structure appearing in each map varies slightly from left to right. The differences in overall mass density are visually apparent, caused by varying $\Omega_\mathrm{M} h^2$. In Cosmo2, each successive plot shows halos from a slab $\sim$100--150 cMpc offset, hence the structure appears mostly distinct between neighboring maps. However, the maps are more statistically similar due to the fixed physical matter density.}
    \label{fig:3cosmos}
\end{figure*}

The HMF for the fiducial $\Omega_{\mathrm{M}} = 0.32$ cosmology, as well as two other variations for each set, are shown in the upper panel of \Cref{fig:3cosmos}. In Cosmo1, the variation of $\Omega_{\mathrm{M}}h^2$ impacts the HMF significantly, causing a steep loss of halos at high masses for the low-$\Omega_{\mathrm{M}}$ universe. The Cosmo2 variations are more subtle, with a similar HMF across all runs. For the mean matter density to be fixed across the Cosmo2 variations, the low-$\Omega_{\mathrm{M}}$ runs must compensate for the loss of high-mass halos by creating more low-mass ones; the trend toward this can be seen in the slope of the HMF curves, but our choice in resolution limit for the simulations prevent this compensation from being clearly evident.

The differences in large-scale structure (LSS) can be seen visually in the lower panels of \Cref{fig:3cosmos}, where we examine the projected halo mass density in a box of fixed comoving volume across all cosmologies, centered at the same redshift in the simulation, $z=0.5$, where $\sigma_8$ has been fixed. The projected map is smoothed with a Gaussian filter of FWHM=10~Mpc in each cosmology, half of the scale at which we will examine the orientation of structure. The redshift $z=0.5$ corresponds to a different $\chi$ for each cosmology (see \Cref{tab:cosmo_var}), so despite the same seed being used for the initial density field in each run, the location of structure in the different maps is distinct (especially in Cosmo2). In Cosmo1, the impact of the steep increase in massive halos toward higher $\Omega_{\mathrm{M}}$ is easily visible, dominating the changes amongst the cosmologies and obscuring variations in the large-scale clustering patterns. However, the Cosmo2 variations are more subtle. Fixing the matter density but varying the Hubble constant and matter-$\Lambda$ interplay across the simulations induces non-trivial changes to the clustering patterns. 
Extended structures are evident in all variations, but the number of structures, overdensities of the regions, and degree of local alignment vary with the cosmology. In the volume shown, there appear to be fewer multi-cluster conglomerations in the $\Omega_{\mathrm{M}}=0.22$ universes, and those that do exist are more filamentary. In the highest $\Omega_{\mathrm{M}}$ universe, the prominent features are overdense on large-scales and there are more isotropic features. For values of $\Omega_{\mathrm{M}}=[0.27, 0.37]$, we see a middle ground between filamentary structures and clumpy overdensities.

Finally, we create $y$-maps from the halos by pasting gas pressure profiles as a function of mass and redshift. This is done with our chosen fiducial model, the pressure profile parametrization from \cite{Battaglia2012b}, which is described in detail in the forthcoming Section~\ref{sec:gas_pasting} (Eq.~\ref{eq:BBPS}). After creating mock realizations of the tSZ $y$ signal, the sky-averaged mean value $\langle y \rangle$ is removed from the maps to mimic the lack of sensitivity to the global monopole in observational data.

In reality, of course, a large fraction of gas at intermediate pressures and temperatures is distributed in the intergalactic medium \citep{Cen1994, Dave2001, McQuinn2016, Cuciti2022-ha} which only full hydrodynamic simulations can hope to represent. However, a halo-based gas pasting approach is a rapid and simple way to generate correlated mock tSZ and galaxy data for comparison to large surveys. Such an approach can reasonably predict tSZ for hot ($\sim10^6-10^8$ K) gas in large (cluster- and large-group-hosting) halos, especially when including simple prescriptions which account for at least some of the effects of baryonic feedback \citep{Battaglia2012b, OsatoNagai2023MNRAS.519.2069O, Greco2015, planck2016tSZ, Horowitz2017MNRAS.469..394H, Gong2019MNRAS.486.4904G}.

Although motivated by self-similarity arguments \citep{Kaiser1986}, it is not entirely clear whether the prescriptions used in this work can be accurately used to create $y$ maps for such broad cosmological parameter variations. Studies of the dependence of cluster scaling relations on cosmology with hydrodynamic simulations have shown the self-similar relations hold up well \citep{singh2020}, but it is possible the extension to lower masses could hold heretofore unknown cosmology dependency. For future efforts, especially when creating a framework for cosmological parameter inference, a robust approach would be to validate the pressure model at all scales with hydrodynamic simulations with jointly varied cosmology and gas physics. However, such simulations, with large enough boxes to create ample amounts of groups and clusters, are only very recently becoming available \citep{flamingo2023}, so this is beyond the scope of this work. 

Having retained only halos with $M>10^{13}$\Msun, the very low-mass halo contributions to $y$ are missing from the mock maps. Based on the level of contribution of this mass range to oriented stacks in hydro simulations \citep{Lokken2023MNRAS.523.1346L}, we expect this to bias the stacked signal downward by $\sim5-10\%$ in an $r$-independent manner, except at very small $r$ where the bias is smaller. This bias is likely to be dependent on the cosmology and gas physics, but expected to be a sub-dominant effect to that of the more massive halos.

\section{Stacking methods} \label{sec:methods}

The oriented stacking technique used in this work closely follows that used and elaborated in greater detail in \cite{Lokken2025ApJ...982..186L}, and is a projected version of the 3D stacking done in \cite{Regaldo2021}. In brief, cutouts are taken from the $y$ map centered on halos in highly superclustered regions, then rotated and stacked along the primary filament axis. 

For the stacking locations, we select massive halos with $M>5\times10^{13}\Msun$. We refer to this sample as the `stacking halos'; the mass range includes galaxy groups to the highest-mass clusters. For observational reference, these correspond roughly to $\lambda>10$ clusters in \redmapper for DES Y1 \citep{McClintock2019}. We isolate those `clusters' within the redshift bin $0.485<z<0.514$, a bin centered on $z=0.5$ which corresponds to 100~Mpc in comoving depth in the fiducial Planck cosmology. We emphasize that the bin is defined in redshift space, rather than comoving space, to imitate the most straightforward procedure to perform observationally.

Next, we reduce the halo sample to include only those embedded in large-scale elongated overdensities (such as superclusters). This criterion is determined using a projected map of the halo mass density. To create this map, we use a $z$ bin twice as wide, also centered on $z=0.5$ ($0.470<z<0.529$, 200~Mpc in the fiducial cosmology). The choice of width is discussed in the context of observational data in \citep{Lokken2022PaperI}, as well as some discussion of projection effects. This map includes all halos in the simulation $M>10^{13}\Msun$. We weight each halo by its mass before adding a point at its (RA, Dec) position to a smoothed 2D \textsc{HEALPix} map \citep{Healpix2005}. The projected halo density map is smoothed with a Gaussian filter with a $\mathrm{FWHM}=35'$, which varies in comoving size from $R=17\,\mathrm{Mpc}\,(\Omega_{\mathrm{M}}=0.22)$ to $R=22\,\mathrm{Mpc}\,(\Omega_{\mathrm{M}}=0.42)$ across the simulations. For a halo $i$ from the full stacking halo sample with R.A. and Dec positions $(\alpha_i, \delta_i)$, we keep only those for which the smoothed map, $F$, satisfies certain constraints. The first, $\nu(\alpha_i, \delta_i)>2$, quantifies the rarity of the overdensity: $\nu=F(\alpha, \delta)/\sigma$ where $\sigma$ is the rms of the map. The second, $e(\alpha_i, \delta_i)>0.3$, quantifies the ellipticity: 
 \begin{equation} \label{eq:e}
     e = \frac{\lambda_1 - \lambda_2}{2 (\lambda_1 + \lambda_2)}.
 \end{equation}
Here, $\lambda_1$ and $\lambda_2$ are the larger and smaller (respectively, in absolute value) of the two eigenvalues of the Hessian matrix, which quantifies the curvature at a given position of the map. The values for these thresholds were explored with different simulations in \cite{Lokken2022PaperI}, and we do not explore other values in this work. Applying these constraints reduces the halo sample by a factor of $\sim16$. The impacts of these constraints on the cosmological sensitivity are discussed in \Cref{sec:enviro_constraints}.

After the sample is reduced, cutouts of side length 4\degree are taken from the $y$-map around each halo. From the halo density map, the eigenvector $\boldsymbol{v_2}$ pointing along the long-axis (the filament direction in a simple straight-filament model) defines the angle by which each cutout is oriented with respect to the R.A. axis, such that the long-axis corresponds to the $x$ axis of the image. Before stacking, we flip each cutout about the $x$ and $y$ axes to enforce that the direction of positive gradient (densifying structure) from the halo density map must be oriented towards the right and upwards ($+x,+y$) in each cutout. Finally, we stack the cutouts by a simple average.

Across the cosmology variations, we conduct the procedure identically in terms of observational quantities (redshift and angular scale), so that in comoving coordinates the procedure differs slightly by the volume in which clusters are selected, the amount of projection in the number density maps, the scale for making $\nu$ and $e$ sub-selections, and the scale defining the orientation.

\section{Analysis of cosmological variations} \label{sec:cosmology_analysis}

\begin{figure*}[t]
    \centering
    \includegraphics[width=\textwidth, trim={0 3cm 0 0},clip]{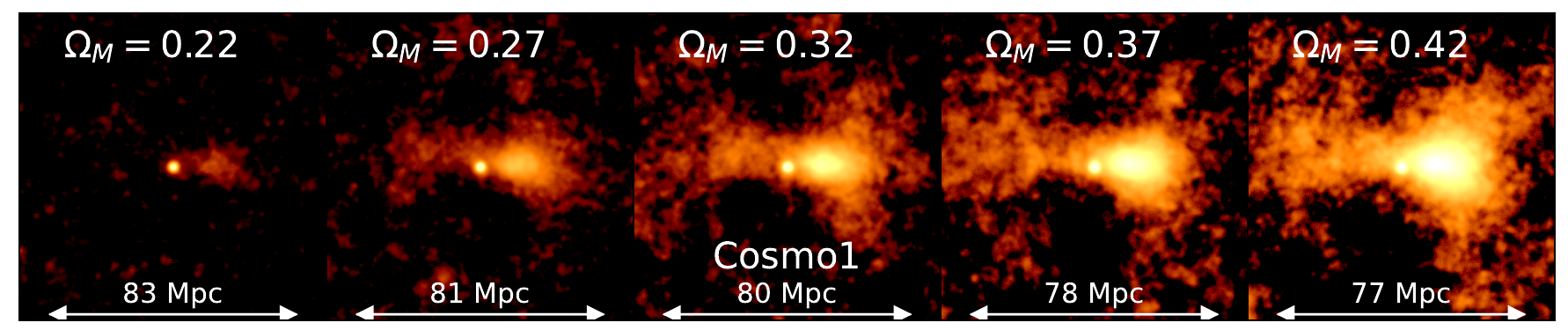}
    \includegraphics[width=\textwidth]{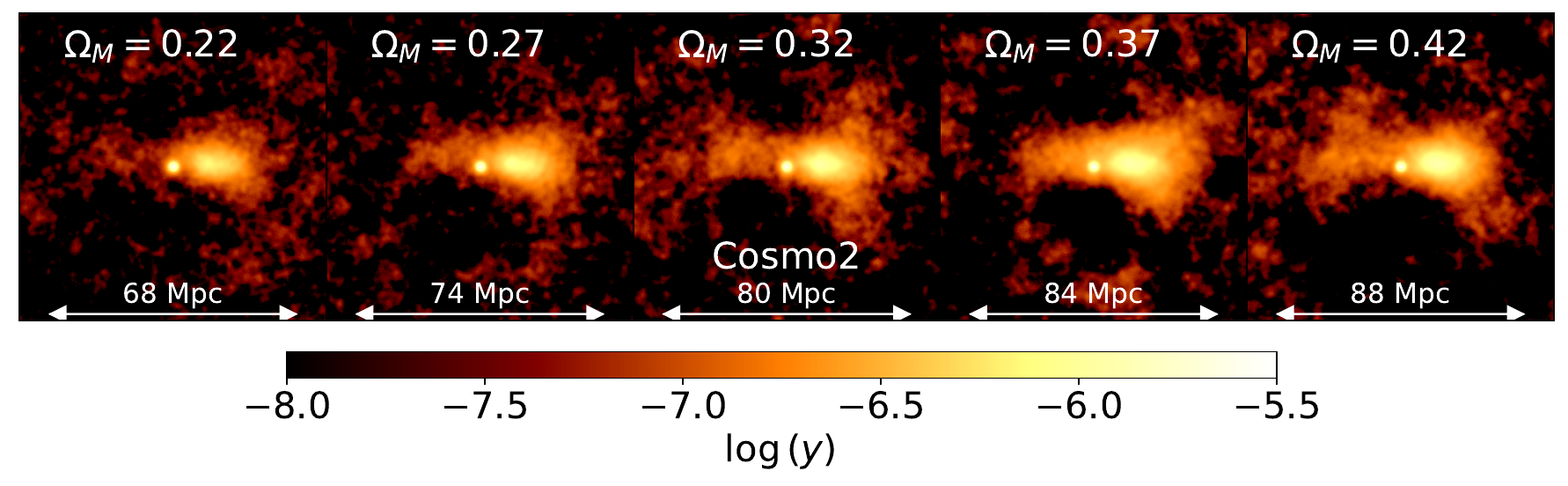}
    \caption{Oriented stacks of Compton-$y$, centered on a sample of $M>5\times10^{13}$~\Msun\ halos constrained by large-scale $\nu$ and $e$ thresholds, for the 5 cosmologies in Cosmo1 (\textbf{above}) and Cosmo2 (\textbf{below}). Plots are shown with a logarithmic color scale to highlight differences in the far-field from the central stacked cluster. The angular size of each image is $2.3^{\circ}\times2.3^{\circ}$ (zoomed-in from the full $4^{\circ}\times4^{\circ}$ stack), but the comoving side-length varies as shown. The $\Omega_\mathrm{M}=0.32$ stack is identical in both rows. The clear horizontal asymmetry about the $y$-axis, and subtler asymmetry about the $x$-axis, is due to the gradient-based flip in the orientation procedure (\Cref{sec:methods}). The images reveal the projected, averaged hot gas from 3D conglomerations of halos located up to comoving distances of 40~Mpc (transverse) and 100~Mpc (line-of-sight) away from the origin in the fiducial cosmology. The signal includes a combination of correlated supercluster structure as well as some uncorrelated foreground/background structure.}
    \label{fig:omegamh2fix_stack_img}
\end{figure*}

\Cref{fig:omegamh2fix_stack_img} shows the stacked images, colored in a logarithmic Compton-$y$ scale, for each set of variations. The stacked clusters at the origin dominate the signal, but the extended structure lying along the horizontal axis is clearly visible in every variation. Due to the gradient-based $+x$ and $+y$ flips, we also see asymmetry about both axes (this was implemented and further discussed in oriented stacks in \citep{Regaldo2021, Lokken2025ApJ...982..186L}, whereas only symmetric orientation was used in L22). In both sets of variations, the amplitude of extended structure as well as the shape and asymmetry vary with $\Omega_{\mathrm{M}}$; the variation is stronger across the Cosmo1 set.

\subsection{Multipole radial profiles} \label{cosmo_analysis}

Next, we compare the first five cosine moments of the stacked images. For each stacked image $I$, these are calculated by
\begin{equation}\label{eq:multipole_moments}
    C_m(r) = \frac{1}{X\pi}\int_0^{2\pi}  I(r,\theta)  \cos{(m \theta)} \dv{\theta},
\end{equation}
where $X=2$ for $m=0$ and $X=1$ for all other $m$, and $\theta$ is the polar angle measured counter-clockwise from the positive $x$-axis. Due to the asymmetric nature of the stacks, there is signal in the analogous sine moments $S_m(r)$ as well, but since it is smaller we focus on the cosine moments in this sub-section and leave the sine moments for \Cref{subsec:integrated_multipole}.

Although we do not run multiple realizations of each variation, we perform a simple estimation of the uncertainties from cosmic variance by dividing each halo sample into 40 sub-samples, where each sample is clustered spatially on the sky\footnote{For the sub-sample division we use the \texttt{kmeans\_radec} algorithm (\url{https://github.com/esheldon/kmeans_radec}).}. Ranging from $\sim5-11^{\circ}$ per side, larger than the 4$^{\circ}$ per side cutouts, we expect stacks from these sub-samples to be mostly (but not completely) independent. We stack each sample, then calculate the covariance between radial bins using each sample of 40 $C_m(r)$ profiles. More statistical completeness is beyond the scope of this work, as this is the first exploration of the cosmological dependence of this technique and we are focused on searching for clear trends in the results.

\begin{figure*} 
    \centering
    \includegraphics[width=.97\textwidth, trim={2.25cm 0 0 0},clip]{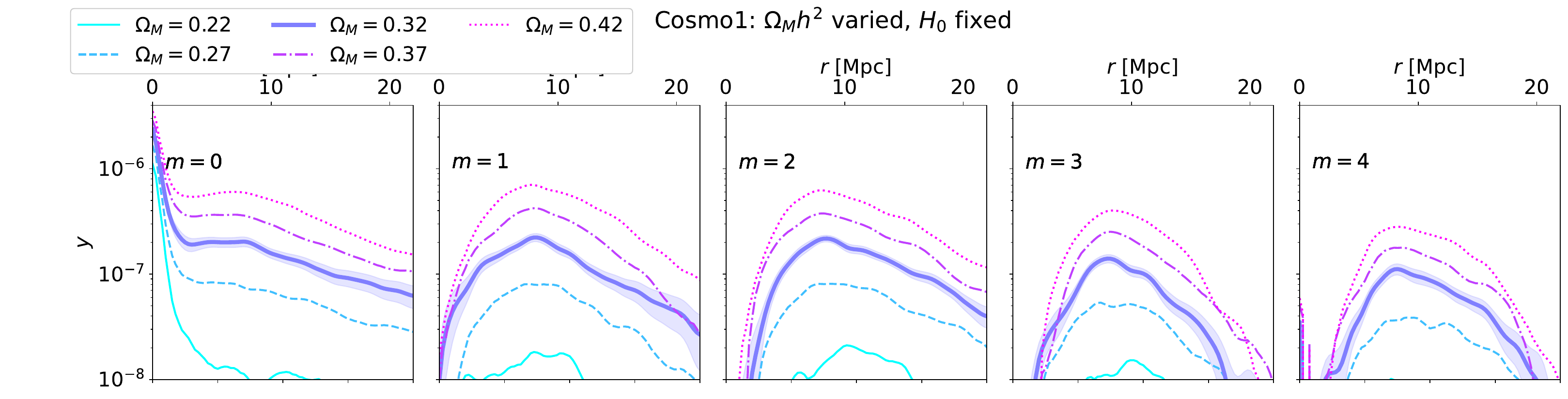} \\
    \hspace*{.18cm}\includegraphics[width=.98\textwidth]{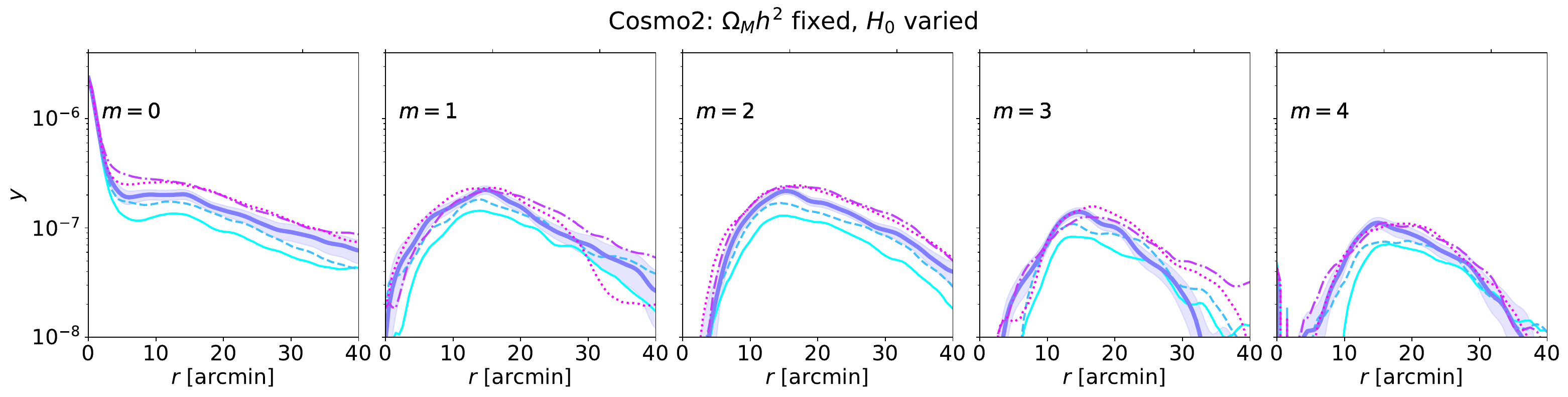}
\caption{Radial profiles of the first several cosine multipoles $C_m(r)$ (\Cref{eq:multipole_moments}) of the oriented stacks shown in \Cref{fig:omegamh2fix_stack_img} for the Cosmo1 (\textbf{upper}) and Cosmo2 (\textbf{lower}) suites. The $x$ axis is the distance from the stack center; the lower axis labels show the angular coordinates, while the upper axis translates this into comoving transverse distance for only the fiducial cosmology. The leftmost plots show the mean-$y$-subtracted $m=0$ curves, representative of the information available in an unoriented stack, while all other plots show higher-order moments, which contain signal due to orientation. The fiducial $\Omega_\mathrm{M}=0.32$ simulation result is shown in bold purple. Variations in $\Omega_\mathrm{M}$ impact the $y$ signal in both the one-halo regime, near $r=0$, and beyond. The effect on Cosmo1 is stronger due to the varying physical matter density; in the one-halo regime the amplitude scales with $\sim\Omega_\mathrm{M}^{1.8}$ and beyond $r\sim4'$ scale more strongly, as $\sim \Omega_\mathrm{M}^{4.5}$. The Cosmo2 variations have a weaker $\Omega_\mathrm{M}$ dependence.}
    \label{fig:m04_omegaM_varied}
\end{figure*}
The results for Cosmo1 and Cosmo2 are shown in the upper and lower panels in \Cref{fig:m04_omegaM_varied}, respectively. The plots are shown in logarithmic scale to highlight the shape and differences in the two-halo regime. In these figures and the following figures in the paper, we only show the profiles out to 40', which corresponds to 23~Mpc in the fiducial cosmology, to focus on the regime where anisotropy is strongest in the stacks. The $m=0$ profiles begin at high values within the clusters, drop rapidly with radius, remain steady until $r\sim7$~Mpc, then slowly fall. The gradual decline is due to the environmental constraints: for an average halo sample, the signal would drop more rapidly to the mean-$y$, but in regions of high large-scale $\nu$, the nearby background is higher than average. The $m>0$ profiles exhibit a rise, peak, and fall; the peak corresponds to the radius at which structure is maximally aligned in the image, which is determined by the smoothing scale for which the orientation is calculated (as discussed in more depth in L22). The 1$\sigma$ errors from sub-sampling are shown only for the fiducial cosmology for visual purposes, but the fractional errors for each cosmology variation are similar. Generally, \Cref{fig:m04_omegaM_varied} shows that the amplitude of $y$ signal in the stacked clusters and the surrounding field depends on the cosmology, with higher $\Omega_{\mathrm{M}}$ (lower $\Omega_\Lambda$) causing a stronger signal as expected. For Cosmo1, which varies $\Omega_{\mathrm{M}}$ with $H_0$ fixed, the majority of differences in the profile strength lie in the one-halo regime ($r\lesssim 2$~Mpc---this is more evident on a linear scale, not shown in any figure). Nevertheless, the two-halo regime also shows a strong dependence on the matter density, represented not only in the isotropic $m=0$ term but also in the higher-order moments.

There is subtler cosmological sensitivity in the Cosmo2 variations (lower panel in \Cref{fig:m04_omegaM_varied}), which exhibits the combined effects of varying the $\Omega_{\mathrm{M}}-\Omega_\Lambda$ split and $H_0$. Here, because of the fixed physical matter density and fixed baryon fraction, the one-halo signal is similar among all variations. Much of the cosmological dependence is in the clustering regime: the amplitude of each profile for both $m=0$ and $m=2$ increases with $\Omega_{\mathrm{M}}$ for $0.22<\Omega_{\mathrm{M}}<0.37$, and a similar trend is seen at the peak of $m=1$. Variations in the profile shapes cause profiles in some moments to cross-over one another. However, the differences in shapes are not significant given the uncertainties.

\subsection{Integrated multipole power} \label{subsec:integrated_multipole}
The differences in profile shape make it difficult to assess cosmological sensitivity in the Cosmo2 stacks. To produce an $r$-independent measure, we evaluate the total multipole power per $m$ using the following equation:
\begin{equation}
    P_m = \int_0^R \left( C_m(r)^2 + S_m(r)^2 \right)~r~dr,
    \label{eq:integrated_power}
\end{equation}
where $C_m(r)$ is from \Cref{eq:multipole_moments} and $S_m(r)$ is calculated with the equivalent equation for sine. $P_m$ is the area integral of the (rotation-invariant) total power in $m$. We evaluate $P_m$ \rr{by integrating over $r$ in angular coordinates (degrees), thus the units for $P_m$ in all figures are deg$^2$.} We choose an integration limit of $R=1$~deg \rr{(34~Mpc in the fiducial cosmology) for all $m$, avoiding the noisy and low-signal outskirts of the signal. However, for some simulations and multipoles, the profiles are still significantly non-zero beyond 1~deg, so not all integrals converge at this limit. In particular, we find that $P_0$ increases 10--50\% across the cosmologies when extending $R$ from 1 to 2 degrees, while the higher-order moments are more converged in general, varying by only a few percent in most cases. Nevertheless, we confirm that the broad trends across cosmology and moment do not depend on the integration limit.}

\begin{figure}
    \centering
    \includegraphics[width=1\linewidth]{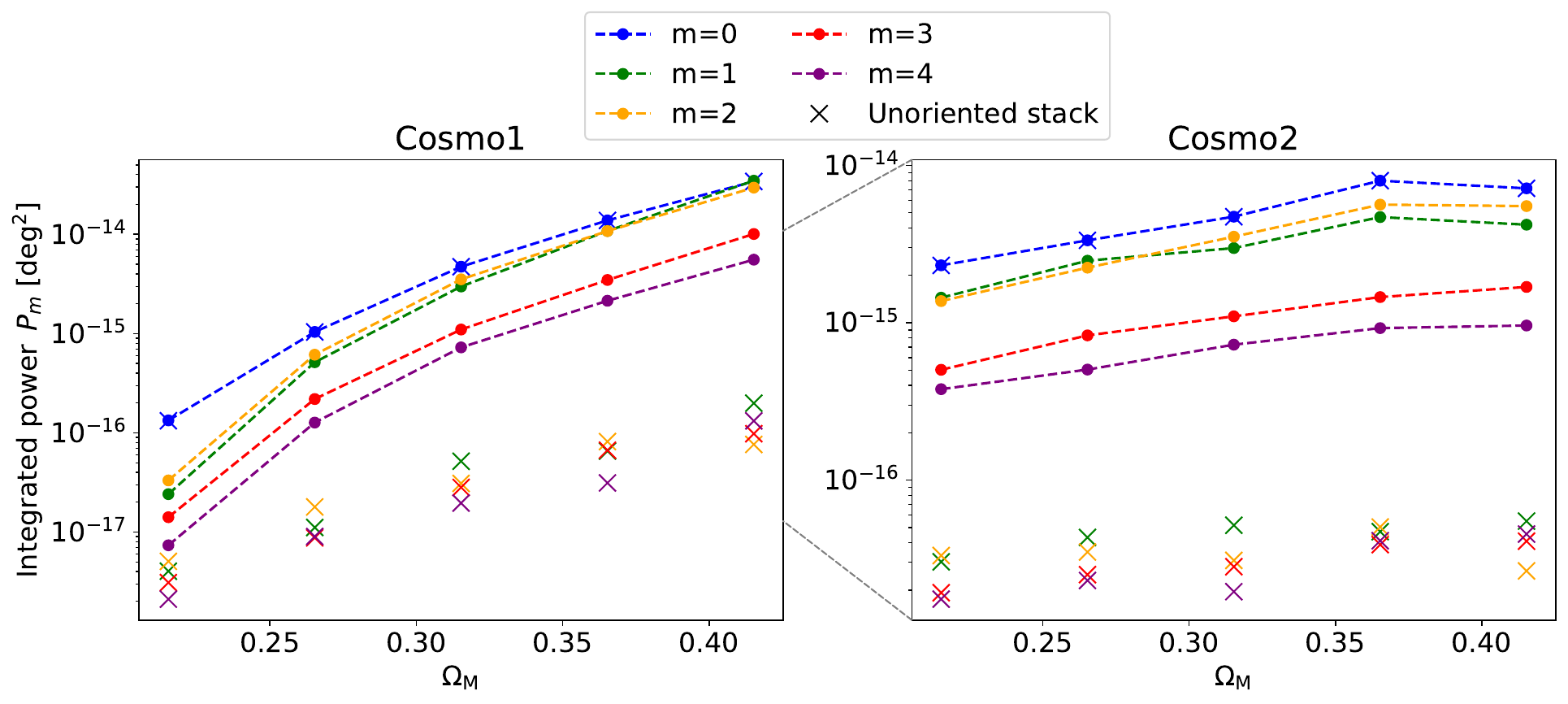}
    \caption{Integrated multipole power for both suites of simulations. The points connected by dashed lines (for visual purposes only) show the power in each moment according to \Cref{eq:integrated_power}. The unconnected `x'-shaped markers show the power for each moment from an unoriented stack for reference; only the $m=0$ moment has significant signal, which is identical to the oriented stack $m=0$ (blue), as expected. Any power in higher-order moments of the unoriented stacks can be thought of as noise power in the corresponding moment from the oriented stack. In both Cosmo1 and Cosmo2, the higher order moments of the oriented stack hold information that could be used to distinguish between the simulations, although the Cosmo2 case is more subtle.}
    \label{fig:integrated_power}
\end{figure}

\Cref{fig:integrated_power} shows the multipole power for the two suites as a function of $\Omega_\mathrm{M}$, in circular markers connected by dashed lines. For all cosmologies, the integrated power in $m=0$ is the strongest; comparatively, the power in $m=1$ and $m=2$ is typically about half, and in $m=3$ and $m=4$ is $\sim10-20\%$.  The dependence on $\Omega_\mathrm{M}$ in Cosmo1 is evident, as before. In Cosmo2, the power is more similar across the suite, but there is still a trend: $P_m$ increases monotonically with $\Omega_\mathrm{M}$ up until $\Omega_\mathrm{M}=0.37$. The power is similar or declines slightly for the highest $\Omega_\mathrm{M}$ value.

To test the level of residual anisotropy in the stacks associated with the finite number of stacked halos, we stack the same halo sample but orient each cutout randomly. In the case of infinite stacked halos, we would expect the stacked signal to be perfectly isotropic as the rotations are uncorrelated with large-scale structure. Therefore, any residual anisotropy in the unoriented stack is due to the finite sample size, and the associated multipole power gives a measure of the noise floor in each moment of the oriented stacks. The randomly-oriented stack power is shown in the same figure in `x'-shaped markers. As expected, the power in $m=0$ is identical. There is negligible power from noise ($\sim1-2\%$) in all other moments. This demonstrates the significance of the power in the $m>0$ moments in the oriented stack. Altogether, \Cref{fig:integrated_power} demonstrates that oriented stacking includes additional cosmologically-dependent information compared to unoriented stacking (or directionless cross-correlations).

\subsection{Source of higher-order moment information}
In the previous section, we demonstrated that the higher-order moments provide distinct information compared to the monopole moment, which ostensibly could help distinguish between cosmologies especially in the two-halo regime. However, without a fine grid of many cosmological variations with which to take derivatives, it is not straightforward to determine which of the varied parameters are the most important in causing these changes.

To narrow down the picture, we can imagine a simple model in which the cosmological dependency is described for the monopole moment, and all higher-order moments scale the monopole with a cosmologically-independent radial profile $A_m(r)$, i.e.:
\begin{equation}
    y_m(r) = A_m(r)~ y_0(r),
\end{equation}
where $y_m$ stands for either $C_m$ or $S_m$.

\begin{figure}
    \centering
    \hspace{2cm}
\includegraphics[width=1\linewidth, trim={0.0 0 2.2cm 0.3cm},clip]{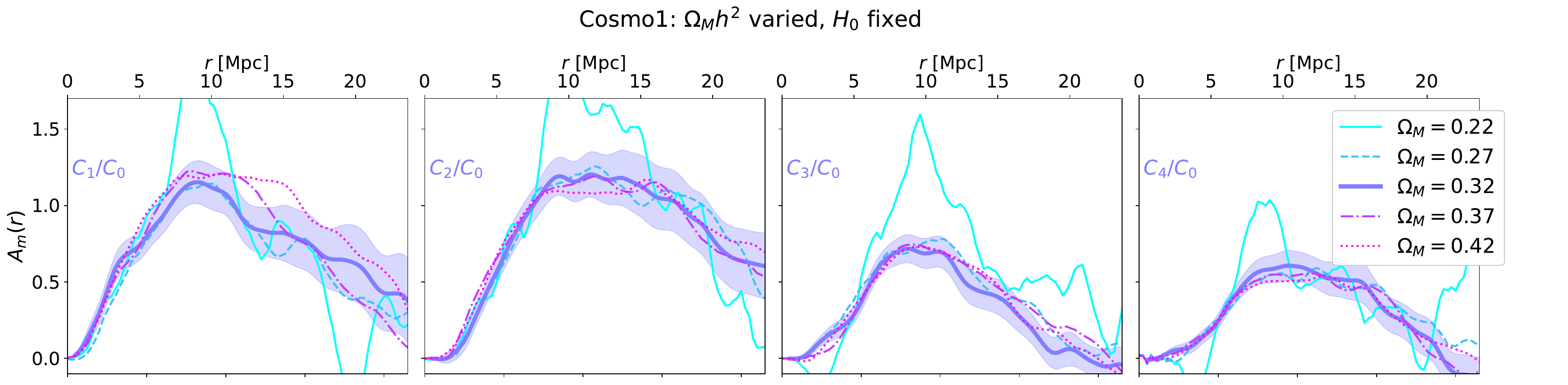}
    \includegraphics[width=.98\linewidth, trim={0.3cm 0 0 0.0},clip]{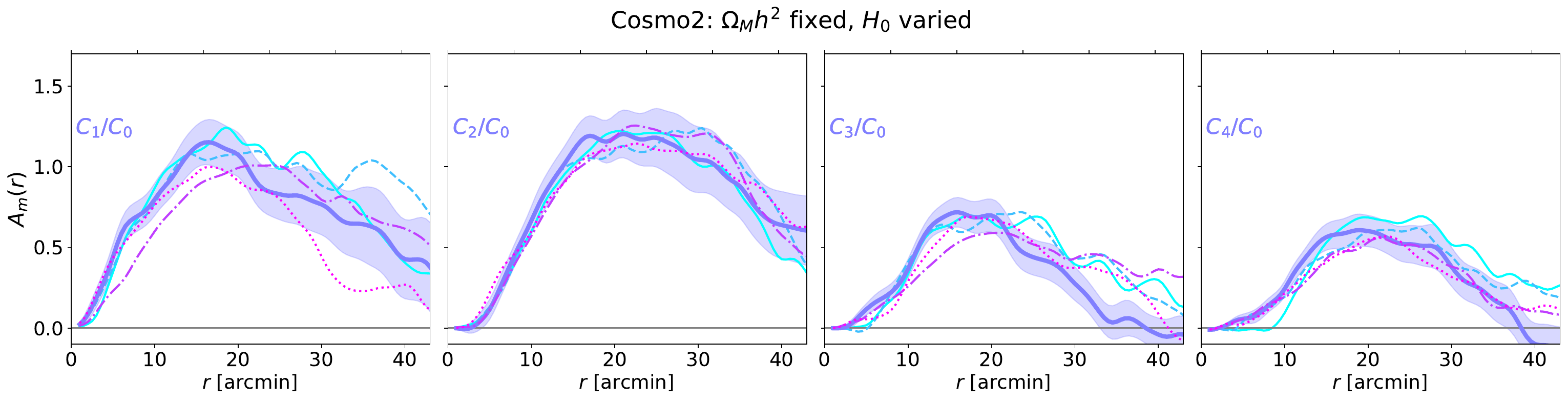}
    \caption{The higher-order moment profiles, normalized by $m=0$, for all variations. The estimated 1$\sigma$ range from cosmic variance, calculated from 30-40 sub-samples, is only shown for the fiducial simulation for visual clarity but are similar or larger in size for the other simulations. \rr{All normalized profiles are consistent within 3$\sigma$ (including $\Omega_\mathrm{M}=0.22$ for Cosmo1, which appears as a visual outlier but is noisier than the rest)}, indicating that the higher-order moments respond to cosmological variations in the same way as the monopole ($m=0$).}
    \label{fig:normalized_plots}
\end{figure}

If this is true, and $A_m$ has no dependence on $\Omega_\mathrm{M}$ or any of the other varied parameters, then measuring $y_m$ simply serves as an independent way to constrain cosmology with sensitivity to the same parameters as $y_0$. We test for this by dividing each $m>0$ curve in \Cref{fig:m04_omegaM_varied} by the corresponding $m=0$ profile for that cosmology run. The results are shown both for Cosmo1 and Cosmo2 in \Cref{fig:normalized_plots}. In both cases, the resulting ratio -- $A_m(r)$ -- is consistent across \rr{all of the simulations within $3\sigma$} (where we have estimated the \rr{cosmic variance uncertainties} of the simulations using \rr{the standard error of the mean over} 40 splits of the data). There are some deviations in the Cosmo2 shapes and for $\Omega_\mathrm{M}=0.22$ in Cosmo1. However, \rr{none are} significant given the errors \rr{(noting that the latter case is the noisiest due to the small halo sample entering the stack, with errorbars larger than those of $\Omega_M=0.32$ shown in the figure)}. Therefore, we do not have strong evidence to conclude that $A_m(r)$ is cosmology-dependent. In other words, the higher-order moments appear to have the same primary cosmology dependence as the $m=0$ moment, and any secondary dependencies (such as the apparent difference in profile shapes in Cosmo2) are too small to demonstrate with simulations of this volume.

If $A_m(r)$ is truly cosmology-independent, there is still value in measuring the oriented, higher-order moments, as they provide additional measurements of the same parameters which affect $m=0$. Observationally, the far-field of an unoriented stack is challenging to measure accurately. Modern CMB experiments are sensitive only to the temperature anisotropies, and thus can measure $\Delta y$, but not the monopole of $y$. The $y$ signal from massive clusters is far above the mean (so $y(\hat{n})\sim\Delta y(\hat{n})$ at cluster position $\hat{n}$), but in the far-outskirts it becomes important to be able to subtract the mean. This is complicated by contamination in the $y$ map; for example, residual long-wavelength primary CMB modes shift the mean $y$ of large patches up or down. Typically the isotropic $y$ profile is estimated by subtracting a stack on random locations \citep{Anbajagane2022}, subtracting the average from an annulus sufficiently far from the stack center \citep{Lokken2025ApJ...982..186L}, or by measuring only \textit{differences} between progressively-large apertures, as in \citep{Schaan2020}. In an oriented stack, the higher-order moments are more straightforward to measure because they are insensitive to the mean, and therefore insensitive to any additive biases that are uncorrelated with large-scale structure. For example, contamination from the primary CMB may have local anisotropy around an individual galaxy cluster, but the contamination averaged over many patches becomes isotropic, even in an oriented stack, because it is uncorrelated with the orientations of late-time structure. This feature makes the $m>0$ moments a valuable addition to the $m=0$ profile, especially for measuring cosmological information from the far-field (clustering) regime.

Having established that the $m>0$ moments add power for measuring the same, or similar, parameters as isotropic profiles, let us discuss the physical origin of this parameter dependence. To do so, we briefly review the literature surrounding isotropic halo-$y$ cross-correlations in the two-halo regime. The cross-correlation signal is presented in \citep{Li2011MNRAS.413.3039L, Fang2012, Vikram2017, Pandey2019} using the halo model. Typically, the 3D halo-pressure cross-correlation is first expressed in Fourier space as a power spectrum, then transformed to configuration space to describe the cross-correlation  between halo $h$ and pressure $P$:
\begin{equation}
\xi^{2h}_{h,P}(r|M) = \int_0^{\infty} \frac{dk}{2\pi^2} k^2 \frac{{\rm sin}(kr)}{kr} P_{h,P}^{2h}(k|M).
\label{eq:xi_twoh}
\end{equation}
The halo-pressure cross-power spectrum $P_{h,P}(k|M)$ describes the power from correlations between the cluster at halo mass $M$ and the pressure contributions from its neighbor halos of masses $M'$, distributed according to the HMF $dn/dM'$. On large scales, it is:
\begin{equation}
P_{h,P}^{2h}(k|M) = b(M) P_{\rm mm}(k)  \int_0^{\infty} dM' \frac{dn}{dM'} b(M') u_P(k|M'),
\label{eq:two-halo_power}
\end{equation}
where $u_P(k|M')$ is the Fourier-transformed pressure profile for halos of mass $M'$ \cite{Hill2018}. This expression assumes that halos are linearly biased by a mass-dependent factor $b(M)$ with respect to the total matter, whose clustering is given by the matter power spectrum $P_{\mathrm{mm}}$. In other words, the tSZ power is related to the matter power by the halo bias of the central halos and the halo-bias-weighted gas pressure of the neighbors. For a stack of halos above a minimum mass $M_{\rm min}$, in our case $5\times10^{13}~\Msun$, \Cref{eq:two-halo_power} must be integrated over all halo masses above this mass, using another factor of the HMF $dn/dM$. The isotropic 2D halo-$y$ cross-correlation (called $C_0(r)$ or $y_0(r)$ in this work) can be described by the integration of the 3D cross-correlation (one-halo plus two-halo) along the line-of-sight \citep[see][for details]{Vikram2017, Hill2018}.

Since we have found that $y_m\sim A(r)y_0$, the higher order moments share the same cosmological parameter dependence. Because each cosmological variation we simulated changes the HMF, and because the pressure profile $u_P(k,M)$ is highly sensitive to mass for the tSZ effect, we infer that the changes to $dn/dM$ are the dominant cause of the differences seen in the oriented stacks.

\subsection{Environmental constraints} \label{sec:enviro_constraints}
The environmental constraints used to sub-select the sample of massive halos, described in \Cref{sec:methods}, also impact the cosmological sensitivity of the measurements. These constraints are motivated, in part, by an effort to measure the highly non-Gaussian late-time universe: in \citep{Lokken2022PaperI}, comparisons with Gaussian random fields demonstrated how enforcing minimum thresholds in the large-scale $\nu$ and $e$ resulted in more highly non-Gaussian halo samples. To examine the impacts these environmental constraints have on cosmological sensitivity, we compare the measurements with and without constraints for the Cosmo2 simulations.

\begin{figure}
    \centering
    \includegraphics[width=1\linewidth]{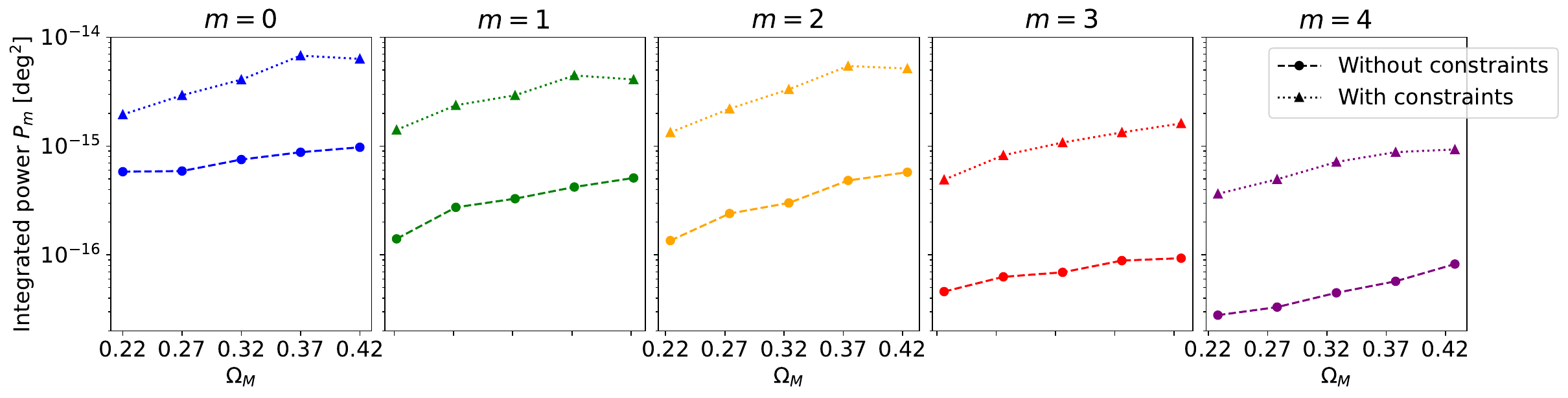}
    \caption{The integrated power across the Cosmo2 variations when the full halo sample is used, rather than the environmentally-constrained sample. Constraints raise the overall power encapsulated in each moment, although they also make the stacks noisier, as discussed in the text. The shape of the cosmological dependency is only slightly impacted.}
    \label{fig:nocuts_power}
\end{figure}

In \Cref{fig:nocuts_power}, we show the integrated power for the Cosmo2 variations for oriented stacks on the full sample of $M>5\times10^{13}$~\Msun\, halos (round points connected by dashed lines). We compare to the stacks after enforcing $\nu>2, e>0.3$ (triangles connected by dotted lines, identical to the right-hand side of \Cref{fig:integrated_power}). With constraints, the overall signal is $\sim4-8\times$ higher. In the unconstrained stacks, there is less relative signal in the higher-order moments compared to $m=0$, while the constrained stacks have a more even distribution of power among the $m\leq3$ moments. The change in power for models with values of $\Omega_\mathrm{M}$ lower and higher than 0.32 is also more drastic when the environmental constraints are placed. This exercise demonstrates how imposing these constraints can, in some cases, increase the cosmological sensitivity.

Although imposing the constraints increases the signal in each multipole and appears to increase the sensitivity to cosmology around the fiducial one, it also depletes the number of stacked objects, and thus yields noisier stacks. For $\nu>2, e>0.3$, the central halo sample after imposing cuts is $12-16\times$ smaller depending on the cosmological model. To properly assess the difference in cosmological sensitivity from the constrained vs. unconstrained samples, one would need simulations with small single-parameter changes and a covariance matrix by which to compute the Fisher information. However, with our existing simulations, we can make a rough comparative assessment by examining the difference in signal with respect to noise.
We assume that the noise per cutout image is uncorrelated with large-scale structure and therefore that noise of the stacked image (and of each multipole profile) scales with $\sqrt{1/N}$. Thus, the noise in the integrated power in multipole $m$ (\Cref{eq:integrated_power}) scales as $1/N$. We compute an estimate of the cosmological sensitivity $S$ from simulation with $\Omega_\mathrm{M}'$ compared to the fiducial cosmology with $\Omega_\mathrm{M}$ by:
\begin{equation}
    S \propto \frac{\mid P_m'-P_m \mid} {\sqrt{(1/N')^2 + (1/N)^2}}.
\end{equation}
We compute the sensitivity scaling for the $\Omega_\mathrm{M}'=0.27$ and $\Omega_\mathrm{M}'=0.37$ run of Cosmo2, and average the two, then compute the ratio of this averaged sensitivity between the unconstrained and constrained sample. By this simple measure, the constrained stacks have 0.9$\times$ the sensitivity of the unconstrained stacks in $m=0$ and 1, 0.8$\times$ in $m=2$, $1.2\times$ in $m=3$ and equal sensitivity in $m=4$.

This exercise shows that the number of objects lost when placing constraints compensates for the increase in cosmological dependence of the signal, and thus examining such regions is not particularly useful for distinguishing between $\Lambda$CDM models. However, the reduction in objects while maintaining similar sensitivity does present an advantage for the constrained case, in that the computational speed of analysis is lower, because it scales with number of objects. Additionally, such constraints may be useful in searching for more exotic cosmologies, such as those that augment non-Gaussianity in particular environments, or include dark-sector physics which only manifests at particular densities.

\section{Sensitivity to gas pressure prescription} \label{sec:gas_pasting}

Unfortunately, the gas pressure profile ($u_P$ in \Cref{eq:two-halo_power}) is not well-constrained except for the highest-mass halos. In order to understand the degeneracies between gas physics and cosmology that will complicate cosmological inference with the superclustering statistics, we create several variations to the tSZ maps for the single \textit{Planck}-cosmology simulation.

The details of the evolution of gas pressure in the cosmic web are complex: for example, gas initially shock-heats as it falls gravitationally into filaments and halos, and later, cooled inner-halo gas is reheated and driven out of halos by active galactic nuclei (AGN) and other astrophysical feedback processes. Various hydrodynamic simulations employ a vast range of prescriptions for these processes and result in varied cosmic web morphologies \citep{Schimd2024A&A...689A.311S}. 

In this section, we only explore variations via spherically-symmetric models of gas \rr{centered on} halos. The range of variations is thus limited, as we do not model non-spherical gas profiles nor \rr{do we model an independent component for diffuse} inter-halo gas; \rr{however, profiles are allowed to extend well beyond $R_{200}$. A previous study of a hydrodynamic simulation suite with strong feedback indicated that modeling the halo-associated gas out to $2R_{200}$ can capture 70-80\% of the oriented-stacked tSZ signal around clusters in the two-halo regime where the present analysis is focused \citep{Lokken2023MNRAS.523.1346L}. Therefore, we expect our modeling (which we extend, in most cases, to $4R_{200}$) to provide a good approximation for a range of gas physics scenarios.} However, gas processes that act in a coherent, anisotropic way with respect to the large-scale structure---\rr{such as anisotropic halo gas aligned with the LSS\footnote{\rr{The gas in clusters is indeed generally aspherical (mostly due to substructure), albeit less elongated than the dark matter \citep{Battaglia2012a}. Observational work has shown that cluster gas does align preferentially along filaments \citep{Baier-Soto2025A&A...704A.228B}}.}}, AGN jets aligned along or perpendicular to filaments, or preferential shock-heating along a particular axis---cannot be captured by our modeling. Some recently-developed semi-analytic models like \citep{Anbajagane2024OJAp....7E.108A, Arico2024A&A...690A.188A} can model anisotropic profiles and/or diffuse particles unassociated with
halos; these would be interesting to test in future work.

Despite their limitations, the halo model variations are not na\"ive, as they are motivated by hydrodynamic simulations. Stronger AGN feedback typically results in gas loss for low-mass clusters and groups, which cannot retain the gas through their gravitational force like massive clusters do. The lower gas pressure alters the $Y-M$ relationship at those lower masses. This also impacts the amount of gas beyond the (somewhat arbitrarily defined) halo `boundaries', as well as its density and temperature; all of the above effects modulate the Compton-$y$ signal \citep[see, e.g.,][]{Yang2022, Sorini2022,  Lokken2022PaperI, Ayromlou2023MNRAS.524.5391A}. We vary features in the pressure profile model to mimic these effects and estimate the possible range of the superclustering observables within the \textit{Planck}-cosmology simulation. We use Websky\footnote{\url{https://github.com/WebSky-CITA/XGPaint.jl}} to paste the projected Compton-$y$ profiles on each halo in our simulations. 
Below, we describe the two models we explore.

\subsection{Fiducial model: BBPS}
Our fiducial model, which was applied to all the cosmology variations in Section~\ref{sec:cosmology_analysis}, is from \citep[][hereafter BBPS]{Battaglia2012b}. Self-similarity predicts a relationship of the integrated Compton-$y$ signal to halo mass of $Y \propto M^{5/3}=M^{1.67}$ \citep{Kaiser1986, Nagai2006ApJ...650..538N}. However, hydrodynamic simulations from BBPS, which incorporated AGN feedback, found a departure from self-similarity. These simulations included halo masses with $M>10^{13}\Msun$, and the authors fit a generalized NFW model \citep{NFW1997} for the thermal pressure profile as a function of $x=r/R_{200c}$ from the halo center, which we adopt:
\begin{align}\label{eq:BBPS}
P_\mathrm{th}^{\mathrm{B12}} &= G M_{200} \,\big[200\,\rho_{\rm cr}(z)\big] \,\frac{f_b}{2R_{200}} \\
&\quad \times P_0 \left(\frac{x}{x_c}\right)^\gamma 
\left[1+\left(\frac{x}{x_c}\right)^\alpha\right]^{-\beta}
\end{align}
where $\alpha=1$ and $\gamma=-0.3$ were both fixed in the study. Cosmology enters through the critical density $\rho_\mathrm{cr}$ and baryon fraction $f_b$. The dimensionless parameters $P_0$, $x_c$ and $\beta$ are all functions of mass and redshift; for each of these parameters generically referred to as $A$, the form of the dependence is
\begin{equation}
    A = A_0\bigg(\frac{M_{200}}{10^{14}\Msun}\bigg)^{\alpha_m} (1+z)^{\alpha_z},
\end{equation} 
where $\alpha_\mathrm{m}$ and $\alpha_z$ describe the scaling with mass and redshift. In BBPS, these parameters were varied to find the best-fit \citep[the values are given in Table 1 of ][]{Battaglia2012b}. The thermal pressure relates to electron pressure by $P_\mathrm{th}=1.932P_e$ for a fully ionized gas, assuming a primordial helium fraction of 0.24. All in all, this model implies a $Y-M$ relation of $Y \propto M^{1.72}$, steeper than the self-similar relation.

Due to the limitations of the simulations, this model was calibrated to galaxy clusters down to a mass of $1.1\times10^{14}\Msun <
M_{200} < 1.7\times10^{14}\Msun$ and out to radius $2R_{200}$. The authors did not attempt to separate the one-halo from two-halo terms, and thus extending the profiles far beyond the one-halo regime is expected to cause double-counting of tSZ contributions. However, cutting the profiles off at $2R_{200}$ will remove the possibility of modeling the type of highly extended gas profiles seen, for example, in the \textsc{Simba} simulations \citep{Dave2019, Sorini2022}. In all gas pasting in \Cref{sec:cosmology_analysis}, we applied the profiles to 4$R_{200}$, as was done for the original Websky simulations \citep{Stein2020}. Now, with the fiducial cosmology simulation, we assess the effect of pasting gas profiles onto halos with different radius cuts. We create several maps with Eq.~\ref{eq:BBPS}, using the fiducial BBPS parameter values, that extrapolate the profile of each halo out to $2R_{200c}$, $4R_{200c}$, and $6R_{200c}$. We apply the model to all halos in the simulation ($M>10^{13}\Msun$).

\subsection{Break model}
In some hydrodynamic simulations, AGN feedback is even more effective at blowing gas out of low-mass clusters and groups than in the BBPS simulations. This causes a steepening in the $Y-M$ relationship \citep{LeBrun2017MNRAS.466.4442L, Yang2022} for those masses. Several observational studies have also shown evidence of this departure from a simple power-law relationship (and thus from BBPS, which did not model these lower-mass halos). In \citep{Hill2018}, the \textit{Planck} $y$-map stacked on SDSS galaxies was found to be best-fit with models for the gas profile that diverged from BBPS through either a mass-dependent scaling in the power law or a break in the power law. The latter is dubbed the ``break model", or sometimes ``uncompensated break model", because it reduces the total thermal energy with respect to BBPS without compensating for the loss. The electron pressure is given by:
\begin{equation} \label{eq:break_model}
    P_e^{\mathrm{B12},\mathrm{br}}(r|M,z) =
    \begin{cases}
        P_e^{B12}(r|M,z), & \text{$M \ge  M_{\mathrm{br}}$}\\
        P_e^{B12}(r|M,z) \left(\frac{M}{M_\mathrm{br}}\right)^{\alpha_\mathrm{m}^{\mathrm{br}}}, & \text{$M < M_{\mathrm{br}}$}\\
    \end{cases}
\end{equation}
In fitting cross-correlation data from an ACT+\textit{Planck} Compton-$y$ map and DES weak lensing maps, \cite{Pandey2021} (hereafter P22) tested this break model. The authors fixed $M_\mathrm{br}$, the ``break mass", to $2\times10^{14}$~\hMsun\ based on the steepening in the $Y-M$ relationship seen in the cosmo-OWLs simulations (with AGN feedback) at roughly that mass \citep{LeBrun2017MNRAS.466.4442L}. Allowing the index \abr to vary, P22 found a preference for a non-zero \abr value in both the \textit{Planck}-only and \textit{Planck}+ACT results. To encompass their results in our simulations, we produce maps for a range of values, \abr$=(0.398,0.972,1.718)$, equal to the (mean-$1\sigma$, mean, mean$+1\sigma$) marginalized constraints found in P22.

The break mass $M_\mathrm{br}$ is not well-constrained by existing data. Among hydrodynamic simulations, it varies: e.g., \citep{Yang2022} studied the $Y-M$ relationship for the \textsc{Illustris TNG} and the \textsc{Simba} simulations, showing that the deviation from self-similarity due to gas depletion from low-mass halos differs between the two in amplitude \textit{and} occurs at different halo masses, primarily because of differing AGN jet feedback implementations \citep{Illustris2018,Dave2019,Sorini2022}. Thus within the framework of the simple power-law models, variations in \Mbr are needed in addition to the \abr variations to be flexible enough to match various hydrodynamical results. Considering this, in addition to the value $M_{\rm{br}}=2\times10^{14}$
\hMsun\ used in P22, we consider a significantly lower value of $M_{\rm{br}}=5\times10^{13}$
\hMsun. We compare these when \abr is fixed to 0.972, leaving joint variations of \abr and \Mbr to future work. All gas physics variations are summarized in Table~\ref{tab:gas_physics_vairations}.

There is another class of model, the compensated break introduced in \cite{Horowitz2017MNRAS.469..394H} and analyzed in \cite{Hill2018}, that we do not implement in this work. This model follows the logic that since gas pushed out from halos must be redistributed to larger radii, the total thermal energy should be conserved. It requires adding a component to the break model defined above, which compensates for the one-halo pressure changes (from varying $M_\mathrm{br}$ and $\alpha_\mathrm{m}^{\mathrm{br}}$) to conserve the global integrated pressure, and thus $\langle y \rangle$. Because this involves adding an isotropic, broad profile when pressure is reduced, we do not expect this model to induce significant differences in the higher-order moments compared to the uncompensated break, and so we leave its exploration for future work.

\begin{table*}[!htbp]
    \centering
    \begin{tabular}{c|c|c|c}
     Model & Parameters fixed    &  Parameter varied & Variations\\
     \hline
     BBPS  & All but radius & $R_\mathrm{max}$ [$R_{200}$] & 2, 4, 6\\
     BBPS + Break & $R_\mathrm{max} = 4R_{200}$, $M_\mathrm{br} = 2\times10^{14}$ & $\alpha_\mathrm{m}^{\mathrm{br}}$ & 0.398, 0.972, 1.718 \\
     BBPS + Break & $R_\mathrm{max} = 4R_{200}$, $\alpha_\mathrm{m}^\mathrm{br}$ = 0.972 & $M_{\rm{br}}$ [\hMsun] & $5\times10^{13},\, 2\times10^{14}$\\
    \end{tabular}
    \caption{Variations to the pressure profile models described in \Cref{sec:gas_pasting}. The rightmost column shows the values we apply for the parameter listed in the `parameter varied' column.}
    \label{tab:gas_physics_vairations}
\end{table*}

\subsection{Analysis}
For each Compton-$y$ map representing an assumption for the tSZ distribution in the \textit{Planck}-cosmology Websky simulation, we perform oriented stacking on mock galaxy clusters with exactly the same methodology as described in \Cref{sec:cosmology_analysis}.

\rr{As alluded to earlier in this section, it is important to recognize the limitations from the fact that the pasted profiles are isotropic. Intuitively, it might seem that variations in the profile amplitudes can \textit{only} impact the $m=0$ moment by construction. However, we do indeed expect to affect higher-order moments with these variations, because we are examining stacked structure well into the two-halo regime. In this regime, the halo mass distribution along the preferred axis of the stacks differs compared to other axes. All of the $y$ models depend on halo mass, and therefore, any simple $y$ amplitude rescaling will change the relative amount of on-axis to off-axis $y$ signal. Thus our simple parametric modeling is useful to understand the realistic impacts to higher-order moments if feedback in the real universe is either locally isotropic or statistically random in direction. However, with isotropic models, what we \textit{cannot} mimic are any feedback effects which are correlated with the large-scale structure orientation.} 

\begin{figure*}
    \centering
    \includegraphics[width=1\linewidth]{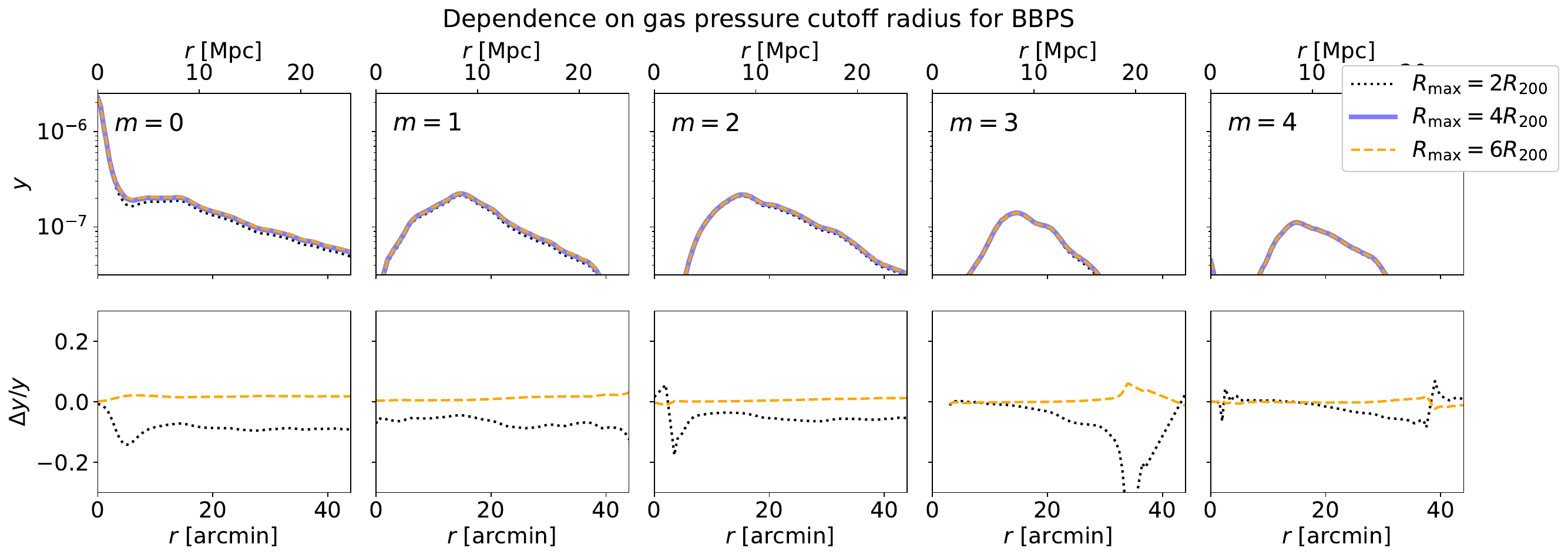}
    \caption{Multipole profiles of stacked $y$-maps with variations in the halo pressure profile cutoff radius within the BBPS model. The blue curve reproduces the results from the fiducial cosmology and gas pressure prescription (with a cutoff at $4R_{200}$) that is shown in \Cref{fig:m04_omegaM_varied} (bolded blue curves in that figure). The lower figures show $C_\mathrm{m,var}/C_\mathrm{m,fid}-1$, the fractional difference of the $2R_{200}$ and $6R_{200}$ models compared to the fiducial $4R_{200}$ \rr{(shown only where $C_\mathrm{m,fid}>10^{-8.5}$ to avoid divergences)}. Decreasing the profiles (black dotted line) has more impact than extending them (yellow dashed), because there is a larger $y$ signal at lower radii.}
    \label{fig:profile_extent}
\end{figure*}

Figure~\ref{fig:profile_extent} shows the impacts of the variations to profile extent within the fiducial BBPS model. The effects can be seen well beyond the average halo radius of the central stacked clusters, because the extended signal is an average over the nearby structure out to a projected comoving distance of 20 Mpc, $\sim10-20\times$ the average central halo radius. As shown in the black dotted curves, limiting the profiles to 2$R_{200}$ subtly reduces the multipole signals by 10-20$\%$ compared to the fiducial 4$R_{200}$ extent. The impacts are similar across multipoles. Extending to $6R_{200}$ (yellow dashed curves) has a small effect, increasing the signal by $<3\%$, due to the rapidly declining pressure profile at these large radii in the BBPS model.

\begin{figure*}
    \centering
    \includegraphics[width=1\textwidth]{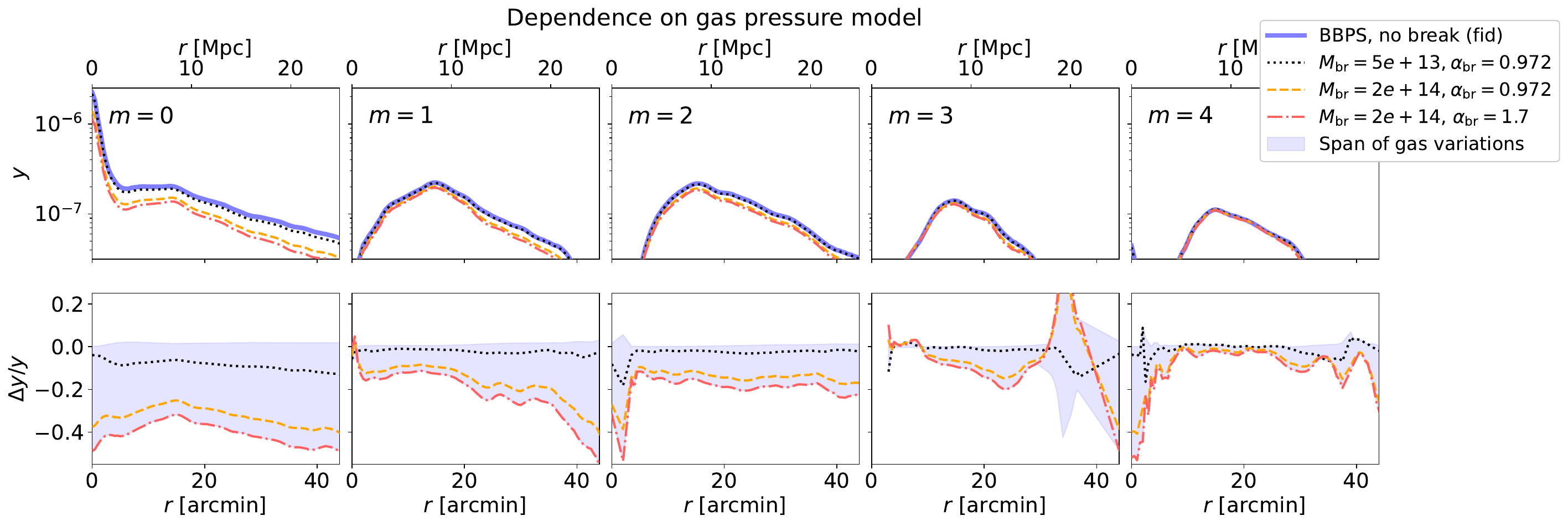}\\
    \caption{Multipole profiles of stacked $y$-maps with variations in the pressure model for halos smaller than a mass threshold. The purple, yellow, and orange curves present variations to the low-mass end of the $Y-M$ relationship using the break model with three different combinations of the values for the mass and power-law modification. Of these, the orange dot-dash curve represents the most extreme suppression of $y$ possible within the $1\sigma$ constraints from \citep{Pandey2022PhRvD.105l3526P}. The shaded purple region spans the minimum to maximum difference induced by the gas model variations explored in this plot and \Cref{fig:profile_extent}; this constitutes a rough estimate of the range within which $\sim68\%$ of gas model trajectories would fall given recent observational constraints. Compared to the range shown here, higher values of $\alpha$ and/or the break mass would suppress the signal further; intermediate and lower values would fill in the space between the existing curves.}
    \label{fig:y_variations}
\end{figure*}

Next, we examine the break-model variations. \Cref{fig:y_variations} shows the BBPS fiducial result reproduced in solid blue, while also displaying the break-model variations in mass cutoff and index in purple, yellow, and orange. Each model is implemented with a cutoff radius of $4R_{200c}$. When the break only affects small masses $M<5\times10^{13}$\Msun, the impacts are less than 10\% in the monopole and $<5\%$ in the higher-order moments. However, when moving the break mass to $2\times10^{14}$\Msun, the impacts are stronger due to the strong slope of the standard $Y-M$ relation. At most, when $\alpha_\mathrm{m}^\mathrm{br}=1.7$, this causes a $\sim45\%$ decrease in $y$ signal. However, it is noteworthy that the higher-order moments are less impacted than $m=0$, which indicates that these variations do not strongly impact the large-scale shape of structure.

In total, we can estimate the range of realistic ($\sim1\sigma$) feedback variations by using the maximum and minimum $y$ values, per $r$ and $m$, as upper and lower bounds. We will represent this span of gas variations in upcoming figures in the following section. The upper bound of this range corresponds, for the most part, to the radially-extended BBPS profile (orange curve in \Cref{fig:profile_extent}), as it was the only model that slightly increases the overall gas pressure in the stacks. Every other variation reduces the pressure, and for most $r$ and $m$ values, the lower bound is given by the most-suppressed break model (red curve in \Cref{fig:y_variations}). We caution that this span is a rough estimate motivated by the current observational literature, but does not account for cases where the feedback effects are anistropic in a manner correlated with the LSS.

\subsection{Degeneracy with cosmology}
\begin{figure*}
    \centering
    \includegraphics[width=0.98\linewidth, trim={0.4 0 .8cm 0},clip]{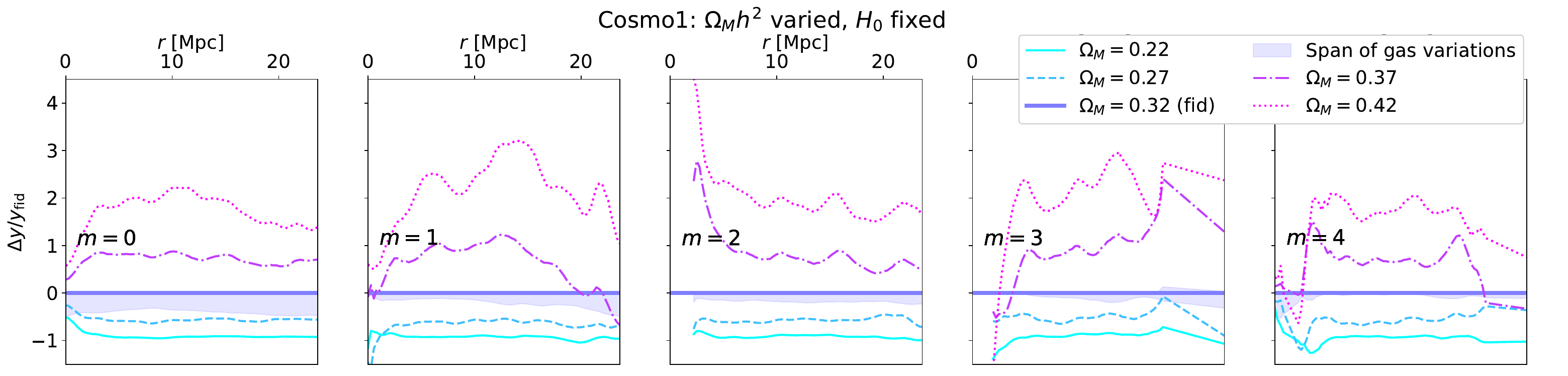}
    \includegraphics[width=0.98\linewidth, trim={0.4 0 0 0},clip]{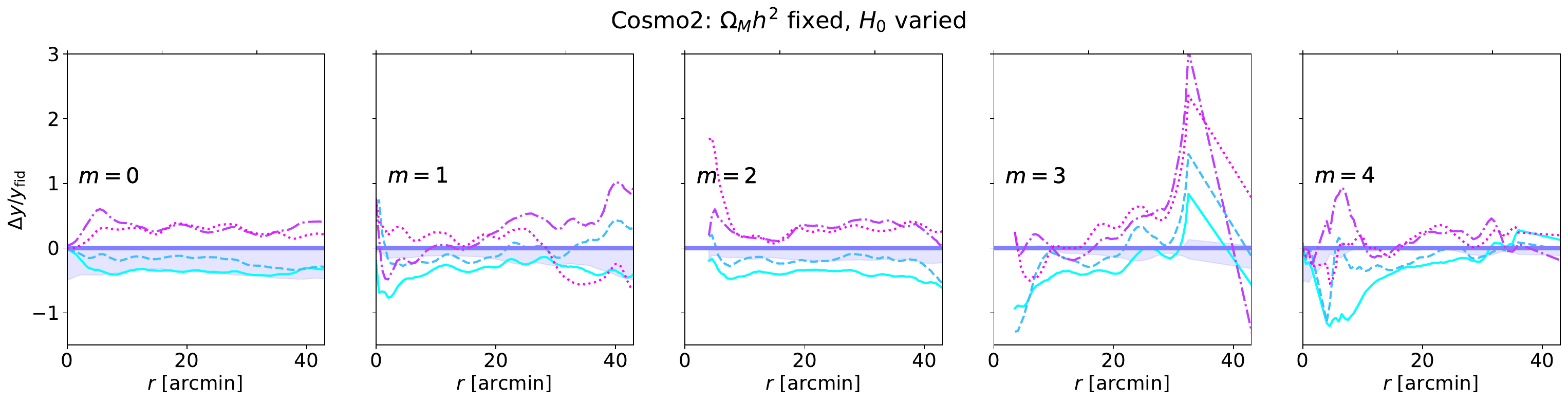}
    \caption{The fractional difference in cosine multipole profiles from the non-fiducial simulations compared to the fiducial one ($C_\mathrm{m,var}/C_\mathrm{m,fid}-1$). The cosmology variations impact the large-scale $y$ signal in a similar manner across multipoles for Cosmo1, while there is more $r$-dependency for Cosmo2. The shaded region is duplicated from \Cref{fig:y_variations}, spanning the full range of gas model variations explored in this work. This range, which can be thought of as the theoretical uncertainty on the fiducial cosmology profiles due to gas pressure modeling, is asymmetrical about $y=0$ because most gas model variations lower the pressure. This gas uncertainty is smaller for the higher-order moments than for $m=0$.}
    \label{fig:pct-diff-como-gas}
\end{figure*}

Having modeled variations in the superclustering signal due to both cosmology and the gas pressure model, it is possible to address the degeneracies between the two. Figure~\ref{fig:pct-diff-como-gas} shows the fractional differences in signal from the varied cosmologies compared to the fiducial one. In the shaded region, we show the span of all gas variations explored for the fiducial cosmology. Although we do not apply gas model variations to the eight other cosmologies, one can expect that the $y$ profiles of the higher-$\Omega_\mathrm{M}$ runs would be more depleted by the break model due to the higher number density of high-mass halos, while the lower-$\Omega_\mathrm{M}$ runs would be less affected. Given the size of the gas range in $m=0$ with respect to the cosmology variations, it is evident our present lack of understanding of gas physics significantly reduces constraining power in $m=0$. However, in both the Cosmo1 and Cosmo2 sets, the higher order moments are less strongly impacted by the gas variations while being similarly sensitive to cosmology as $m=0$. 

\begin{figure*}
    \centering
    \includegraphics[width=0.51\linewidth, trim={0 0 0cm 0},clip]{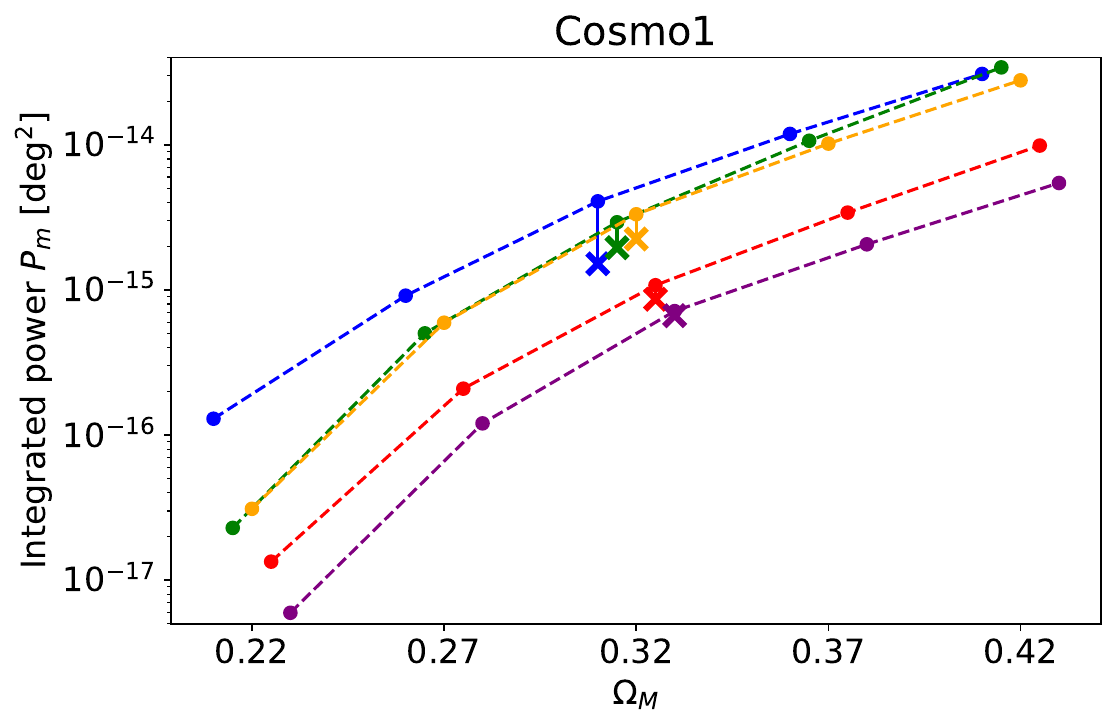}
    \includegraphics[width=0.44\linewidth, trim={1.8cm 0 0 0},clip]{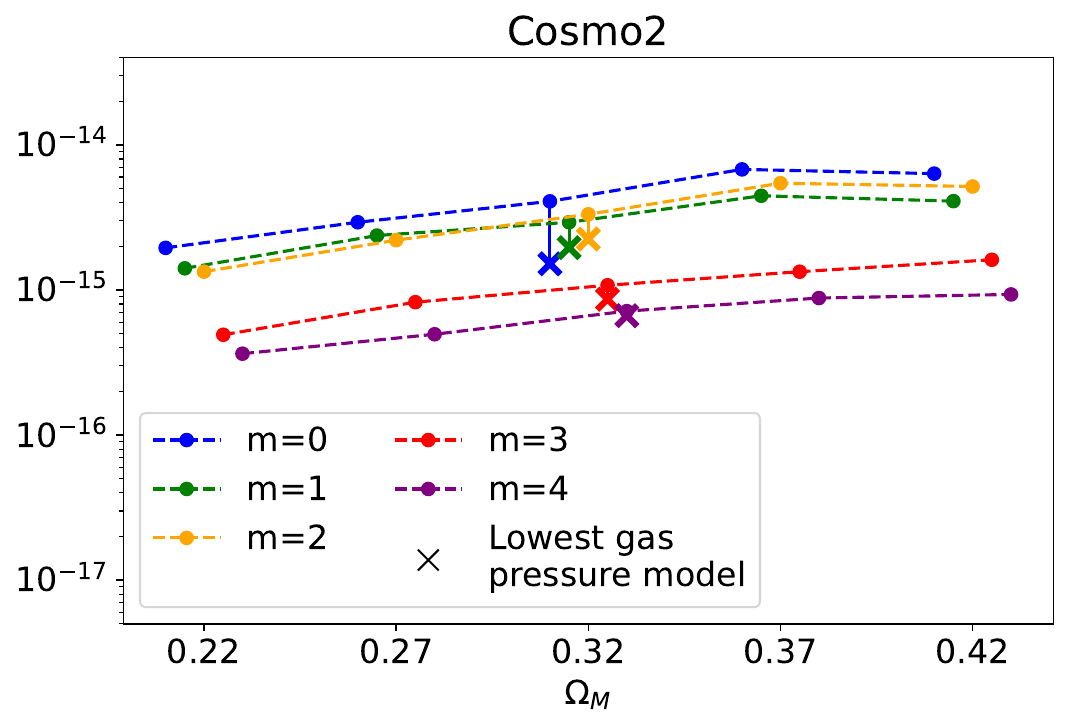}
    \caption{Integrated power in each moment $m$ for each variation defined by $\Omega_\mathrm{M}$. Point markers indicate use of the fiducial BBPS gas pasting. `X'-shaped markers represent integrated power in the break model for the fiducial cosmology, the gas model with most pressure suppression, while the drop-down line indicates the range along which the other gas variations would lie. Points on the $x$-axis are artificially offset for each moment to aid visual distinction. $m=0$ is most sensitive to the gas model variations, while the higher-order moments are progressively less impacted, while retaining cosmological sensitivity. This suggests that the anisotropic moments could be useful for breaking degeneracies between cosmology and gas physics.}
    \label{fig:integrated_power_wfeedback}
\end{figure*}

The same concept is shown through the integrated power in \Cref{fig:integrated_power_wfeedback}, where we repeat the dashed cosmology-dependence lines from \Cref{fig:integrated_power} but add the span of gas variations as a drop-down line. The drop-down line starts from the power (circular marker) in the fiducial Battaglia model stack and terminates in the power (`x' marker) from the most extreme break-model stack, spanning the intermediate variations which lie between these two points. This figure relays that the higher-order power in the fiducial cosmology stack is less impacted by feedback variations compared to the isotropic power. Meanwhile, the difference in power as a function of $\Omega_\mathrm{M}$ is similar across moments. Thus, feedback variations appear to primarily impact the \textit{relative} distribution of power among moments, while cosmology variations rescale power in all moments, maintaining a more consistent distribution across moments. This indicates that, for cosmological inference with oriented tSZ stacks, including higher-order moments could help to break degeneracies with the gas parameters.

In summary, while the current uncertainty in the gas pressure distribution reduces the cosmological constraining power of the superclustering statistics, the impacts of different gas models appear to be distinct from those of cosmological variations as a function of moment $m$. This indicates that using at least some of the higher-order moments will be useful for performing cosmological inference from these superclustering statistics, as they can break the degeneracy between gas and cosmology parameters. \rr{This finding is based on simple modeling that assumes any anisotropies in individual gas profiles are statistically uncorrelated with the LSS orientation; it remains to be seen whether it would generalize to more complex models.}

\section{Conclusions} \label{sec:conclusion}
This work sought to test the potential of constrained, oriented measurements of superclustering in tSZ maps as a cosmological probe. Using the Websky algorithms, we produced two sets of cosmological variations, one in which the dominant effect is the varied physical matter density $\Omega_\mathrm{M}h^2$, and the other which varies the matter-$\Lambda$ ratio and $H_0$ while holding fixed the physical matter density. The set of parameter values was widely-spaced, extending well beyond the regime of current constraints to enable a clear demonstration of the sensitivity of the superclustering statistics. We evaluated the simulations only at $z^*=0.5$, where $\sigma_8(z^*)$ was set to be equal in all simulations to control for the strong tSZ dependence on that parameter.

The cosmological variations alone yielded several key insights:
\begin{itemize}
    \item In both suites, we find cosmological sensitivity in the two-halo / clustering regime.
    \item This cosmological sensitivity persists in the higher-order moments.
    \item For Cosmo1, the parameter variations cause a roughly $r$-independent and moment-independent amplitude scaling in the clustering regime; in other words, higher-order moments ($m>0$) of the stacks have similar parameter sensitivity as $m=0$. This is also true for Cosmo2 in some moments.
    \item Across Cosmo2, there are subtle hints of cosmological dependence of the profile shape with $r$ in higher-order moments, but larger simulations samples are needed to confirm them.
\end{itemize}
In combination, this is a promising demonstration of how locally-anisotropic measures which extend into the far-field beyond massive halos can add significant cosmological constraining power. Simulations in the Cosmo2 suite would be challenging to distinguish by only studying unoriented stacks within the one-halo regime (the usual focus of $y$-stacking studies). Simply extending isotropic measurements into the two-halo regime is useful, but faces observational complications: modern CMB telescopes do not measure the true mean-$y$ value, and long-wavelength fluctuations due to primary CMB contamination make it challenging to accurately zero-out the tail end of the profiles to measure the small-amplitude signals in the far-field. However, the $m>0$ profiles of an oriented stack are more straightforward to measure in this regime because they do not depend on the mean by nature. Therefore, extended profiles from oriented stacking are especially useful for distinguishing between cosmological models such as those in Cosmo2.

We also studied how uncertainties in gas physics modeling may complicate efforts to measure cosmology. We applied several prescriptions for gas pressure to the fiducial simulation using the halo-model framework. These included both the standard BBPS pressure profiles with varied profile extents and an extension to this model which includes a broken power-law, depleting pressure in lower-mass halos to mimic some of the effects of strong AGN feedback. We found that the broken power-law has a much larger impact than varying the pressure profile extent within the BBPS model. Varying the broken power-law exponent versus the mass at which the relationship is broken has similar impacts on large scales, removing pressure from the (integrated and binned) LSS in all multipoles. Comparing the effects with the cosmological variations revealed the following points:
\begin{itemize}
    \item The gas physics variations cause a roughly $r$-independent amplitude scaling in the $m=0$ profiles, similar to the impacts of cosmology variations, indicating near-degeneracy between cosmology and gas physics.
    \item However, the gas pressure variations have less impact on the higher-order moments than $m=0$, whereas the cosmology variations have similar impacts regardless of moment, indicating that analyzing oriented stacks using more than one moment can help to break the gas-cosmology degeneracy.
\end{itemize}
Both sets of conclusions indicate that it would be beneficial for the future of the subject to include orientation, and analysis of anisotropic information, in stacked profiles. However, an important caveat to our analysis is that we performed all gas variations using isotropic halo painting, limiting the space of possibilities covered. In a realistic hydrodynamical simulation, the higher order moments of the tSZ profile may depend on directional details of the feedback implementation---such as coherent (mis)alignment of AGN jets with large-scale structure---which could \rr{mimic cosmological signatures in higher $m$ and thus} increase the degeneracy with cosmology.

Several avenues would be interesting to pursue for future simulation-based work. Further work with hydrodynamical simulations including feedback variations is necessary to verify the degeneracy-breaking power of the higher-order moments. The superclustering statistics used in this paper may be also useful for probing more exotic cosmological models beyond $\Lambda$CDM, such as particular types of primordial non-Gaussianity which induce additional local anisotropies in structure, or couplings in the dark matter-dark energy field which could impact the triaxial growth of structures like filaments and superclusters.  Ultimately, robust inference with these statistics will require predictions that jointly vary cosmology and astrophysics parameters. Such predictions may be possible in the future with advancements made in semi-analytic modeling, emulators, and/or simulation-based inference.

Observationally, the technique used in this work has been thus far demonstrated using maps of the tSZ, kSZ, galaxy weak lensing, and galaxy number density \citep{Lokken2022PaperI, Lokken2025ApJ...982..186L, Hadzhiyska2025PhRvD.111b3534H} in combination with photometric galaxy data. For a point of comparison, in \cite{Lokken2025ApJ...982..186L}, the $m=2$ moment of an oriented stack of ACT $y$ on $\sim800$ DES clusters was measured at 7$\sigma$ (taking into account only statistical uncertainty). We can use the data point at the peak of this $m=2$ profile as a reference, which was measured at $y\sim1\times10^{-7}$ with 25\% uncertainty. (The signal value, while the same order of magnitude as the profiles shown in this work, should not be compared because of differences in the methodology and samples.) If the fiducial cosmology+gas profile in \Cref{fig:m04_omegaM_varied} had the same percentage uncertainty, it would be possible to distinguish between the peak data points in neighboring Cosmo1 variations, which are separated by 2-3$\times$ the size of the errorbars. However, it would not be possible to distinguish between the more subtle Cosmo2 variations. Although this rough comparison provides some context, a full forecast of the level of cosmological constraints from current and future tSZ maps would require accounting for the full profiles, producing corresponding mock covariance matrices with all known sources of noise, and considering systematic errors such as CIB contamination, which is beyond the scope of this work.

Generally, statistical errors decrease with larger stacking samples. The stacking samples can be augmented by using wider redshift bins, although this is likely to lose cosmological constraining power related to redshift evolution. More objects can be stacked by using a lower-mass sample, but this also decreases the signal. Greater overlap between CMB surveys and galaxy surveys will be helpful to increase sky area and thus increase stack sample sizes. The full southern sky coverage from the now-operating Vera Rubin Observatory \cite{LSST_Science_book2019} will increase galaxy overlap with ground-based CMB telescopes, and errors will also decrease with reduced noise in maps from the Simons Observatory \cite{SO2019}. The cosmological sensitivity demonstrated in this work and the future improvements in data make the oriented stacking technique a promising avenue for future cosmological inference.

\section*{Acknowledgments}

Canadian co-authors acknowledge support from the Natural Sciences and Engineering Research Council of Canada. Websky computations were performed on the Niagara supercomputer at the SciNet HPC Consortium. SciNet is funded by: the Canada Foundation for Innovation; the Government of Ontario; Ontario Research Fund - Research Excellence; and the University of Toronto.

IFAE is partially funded by the CERCA program of the Generalitat de Catalunya.

ML acknowledges financial support from the Spanish Ministry of Science and Innovation (MICINN) through the Spanish State Research Agency, under Severo Ochoa Centres of Excellence Programme 2025-2029 (CEX2024001442-S).

In the course of preparing this paper, several artificial intelligence (AI) tools were used to assist with coding and editing. The GitHub Copilot plugin for Visual Studio Code was installed during part ($\sim50\%$) of the programming process, providing code-completion suggestions, some of which were implemented. All code suggestions were reviewed and modified as needed by the first author. ChatGPT and the Writefull installation on Overleaf were used to provide editing suggestions for grammar, typos, and text consolidation, some of which were implemented to enhance readability.

\bibliographystyle{JHEP.bst}

\bibliography{websky_supercluster}

\end{document}